\documentclass[12pt]{article}

\usepackage{epsf,epsfig,amstext,amsfonts, amsmath,latexsym,rotating}
\usepackage{graphicx}
\usepackage{cite}

\def\be{\begin{equation}}
\def\ee{\end{equation}}
\def\ba{\begin{eqnarray}}
\def\ea{\end{eqnarray}}

\def\bq{\begin{quote}}
\def\eq{\end{quote}}

\newcommand{\beq}{\begin{equation}}
\newcommand{\eeq}{\end{equation}}
\newcommand{\beqa}{\begin{eqnarray}}
\newcommand{\eeqa}{\end{eqnarray}}

\newcommand{\lmk}{\left(}
\newcommand{\rmk}{\right)}
\newcommand{\lkk}{\left[}
\newcommand{\rkk}{\right]}

\newcommand{\te}[1]{\text{#1}}

\newlength{\shift}
\newlength{\cwl}
\newlength{\cws}
\newlength{\cw}
\newlength{\cwh}
\newcommand{\pb}[2]{\parbox{#1}{#2}}

\newcommand{\bc}{{\tt BraneCode}{\small $^{\mbox{\textregistered}} $}}

\def\ltap{\ \raise.3ex\hbox{$<$\kern-.75em\lower1ex\hbox{$\sim$}}\ }
\def\gtap{\ \raise.3ex\hbox{$>$\kern-.75em\lower1ex\hbox{$\sim$}}\ }
\def\gl{\ \raise.5ex\hbox{$>$}\kern-.8em\lower.5ex\hbox{$<$}\ }
\def\roughly#1{\raise.3ex\hbox{$#1$\kern-.75em\lower1ex\hbox{$\sim$}}}

\begin{document}
\thispagestyle{empty}
\begin{flushright}
{\tt hep-th/0309001} \\ CITA-2003-30
\end{flushright}
\vspace*{0.2cm}
\begin{center}
{\Large \bf Braneworld Dynamics with the } {\Large \bc}
\\
\vspace*{1cm}
{\large Johannes Martin, Gary N. Felder, Andrei V. Frolov,

Marco Peloso, Lev A. Kofman}

\vspace{0.5cm}

{\em Canadian Institute for Theoretical Astrophysics}\\
{\em University of Toronto}\\
{\em 60 St.~George St., Toronto, M5S 3H8, ON, Canada}
\vspace{0.2cm}

{\tt johannes@physics.utoronto.ca, felder@cita.utoronto.ca,
  frolov@cita.utoronto.ca, peloso@cita.utoronto.ca,
 kofman@cita.utoronto.ca}

\vspace{1cm}
ABSTRACT
\end{center}
We give a full nonlinear numerical treatment of time-dependent 5d
braneworld geometry, which is determined self-consistently by
potentials for the scalar field in the bulk and at two orbifold
branes, supplemented by boundary conditions at the branes.  We
describe the \bc, an algorithm which we designed to solve the
dynamical equations numerically.  We applied the {\tt BraneCode} to
braneworld models and found several novel phenomena of the brane
dynamics.\\
Starting with static warped geometry with de Sitter branes, we found
numerically that this configuration is often unstable due to a
tachyonic mass of the radion during inflation.  If the model admits
other static configurations with lower values of de Sitter curvature,
this effect causes a violent re-structuring towards them, flattening
the branes, which appears as a lowering of the 4d effective
cosmological constant.\\
Braneworld dynamics can often lead to brane collisions.  We found that
in the presence of the bulk scalar field, the 5d geometry between
colliding branes approaches a universal, homogeneous, anisotropic
strong gravity Kasner-like asymptotic, irrespective of the bulk/brane
potentials.  The Kasner indices of the brane directions are equal to
each other but different from that of the extra dimension.  \vfill
\setcounter{page}{0}
\newpage

\tableofcontents
\newpage

\section{Introduction}

Braneworlds embedded in higher dimensions bring new powerful concepts
in cosmology \cite{rev}, as well as they do in
fundamental superstring/M theories and phenomenological  high energy  particle
physics \cite{RS, HW, Lukas, RS1}.
Branes enrich our view with new ideas underlying the four
dimensional effective field theory, bringing, most importantly, new
geometrical images beyond it. For instance, in this context the four
dimensional cosmological constant is the curvature of the brane, and
we should explain why the brane we live in is almost flat. Inflation,
by contrast, corresponds to curved branes. There are interesting ideas
for realizing early universe inflation in braneworld scenarios where,
for example, concepts such as the inflaton potential and inflaton decay are
reformulated in terms of brane-brane or brane-antibrane interactions
\cite{infl}, or topics of  brane collisions.

 The language and images of the braneworld theories are commonly
 shared by fundamental and phenomenological
high energy physics theories, general relativity
and brane  cosmology, with different degrees of trade
between dynamics and  simplification.

The compactification of the extra space is often a key issue in brane
models. For example, a stable radion field, controlling the volume of
the extra space, is usually needed to recover standard four
dimensional cosmology at ``late'' times, and to fulfill precision
tests of general relativity. In addition, the compactification has to
be consistent with the fact that bulk fields have not yet been excited
in accelerator experiments, because they are too massive and/or too weakly
coupled to the visible brane. Schemes for compact inner dimensions
typically rely on the interplay between bulk and brane dynamics.

So far, the control of dynamical, time-dependent, cosmologically
relevant solutions in the fundamental, comprehensive theory has been
rather limited. Relatively simple, yet meaningful, are the five
dimensional phenomenological braneworld models with two orbifold
branes at the edges, where our $3+1$ dimensional spacetime is one of
the branes embedded in the (warped) five dimensional space.  These
models often include one or more bulk scalar field(s) $\phi$ with the
potential $V(\phi)$ and self-interaction potentials $U_i(\phi)$ at the
two branes, as well as other fields $\chi$ localized at the branes.
This class of braneworld models covers many interesting constructions
including the Ho\v{r}ava-Witten theory \cite{HW}, the Randall-Sundrum
model with a phenomenological stabilization of branes \cite{gw,kk},
warped geometry with bulk scalars \cite{super,Dewolfe}, supergravity
with domain walls \cite{cr}, and others.

There is a number of important papers studying static geometries with
branes, including flat stabilized branes, in agreement with low energy
physics, curved de Sitter branes, corresponding to early universe
inflation, and small fluctuations around static warped geometries.
The cosmological evolution has been studied in some of these
pioneering works in the simplest cases in the absence of any scalar
field \cite{BDL,SMS}.  The 4d evolution on the brane, in terms of
effective Friedmann equations, is typically different from the
standard four dimensional cosmology.  The effective 4d Einstein
equations on the brane were also derived for the more general
situation of self-consistent geometry with the bulk/brane scalar field
\cite{MW}.

Standard cosmology can be recovered after the extra dimensions have
been stabilized \cite{CGRT}.  In this respect, the presence of bulk
scalar field(s) becomes crucial. In this more relevant case, however,
the evolution is only known for limiting situations. In general, the
system is very complicated, since the effective four dimensional
Einstein equations are not closed and require solutions of the full
five dimensional equations \cite{L}.

In this paper we address the problem of self-consistent fully
nonlinear dynamics of the 5d braneworld with bulk scalar field(s) with
a bulk potential as well as brane potentials required for brane
stabilization. We consider plane-parallel orbifold branes, so that the
problem is effectively two dimensional, with the metric and the fields
depending on time and the extra dimension $(t, y)$.  Although this
setting is already too involved to be studied analytically, it is
still significantly simpler than what we will need in order to
understand cosmological solutions in ``realistic'' higher dimensional
theories. However, as we shall see, already this step requires the
introduction of new techniques. We have designed and used a numerical
code to solve the partial differential equations describing the system
of non-linear gravity and a scalar field, complementing the existing
approaches to this problem found in the literature.

We aim for generic features of braneworld dynamics, in particular,
attractor solutions.  They will generally depend on the specific
braneworld model, i.e.\ on the bulk/brane scalar field potentials
$V(\phi), U_i(\phi)$.  As a simple illustration, consider a static
five-dimensional warped geometry with a bulk scalar and four
dimensional slices of constant curvature. It is possible to exhaust
the global properties of static warped geometry using the method of
phase trajectories \cite{FFK}, although some details of the phase
portraits depend on the bulk potential. For this problem the phase
space is three dimensional, the critical points (like attractors,
repulsors, and others) can be identified, and all trajectories
(solutions) start and end at critical points.

The $(t, y)$ problem of the time-dependent braneworld dynamics is much
more complicated than the static $(y)$ problem. Using the {\tt
  BraneCode} we were trying to give examples of interesting dynamical
features.  We notice several novel phenomena including a transition
between different warped states and a generic strong gravity solution
of colliding branes.

The plan of the paper is the following.

In Section~\ref{sec:setup}, we give the setup of the braneworld models
and write down the bulk equations supplemented by the junction
conditions at the branes. We pay especially close attention to the
choice of the gauge in order to have a suitable metric for the
numerical calculations. It turns out that, without any loss of
generality, it is possible to choose coordinates where the two branes
have fixed positions along the fifth direction $y$. The geometry is
described by two metric components $A(t, y)$ and $B(t, y)$.

In Section~\ref{sec:numerics}, we describe the {\tt BraneCode}, an
algorithm we use to solve the dynamical equations numerically. At the
moment we have slightly different implementations of the {\tt
  BraneCode} (in $C++$ and Fortran$-90$) in order to cross-check them.
We plan to release the {\tt BraneCode} using the most optimized and
documented version (in $C++$). As is typical for numerical GR
problems, we have to take initial conditions for the metric and fields
which satisfy the constraint equations at an initial time
hypersurface. In Subsection~\ref{sec:static}, we discuss how to
fix the initial conditions for the numerical integration with the {\tt
  BraneCode}.

In Sections~\ref{sec:rs}--\ref{sec:boom}, we apply our {\tt BraneCode}
to three braneworld models where we encounter qualitatively different
dynamics.

In order to check our numerical code, in Section~\ref{sec:rs} we first
apply it to a simpler brane model without a scalar field, for which
analytic solutions are known.  As a playground here we use the
Randall-Sundrum model of two branes embedded in an AdS 5d background.
In Subsection~\ref{sec:rs1} we first use the static RS solution
without moving branes.  In Subsection~\ref{sec:ads-schw} we extend the
calculations to the case of moving branes. In this case the 5d
geometry is described by the analytic AdS-Schwarzschild solution (with
the mass of the virtual 5d black-hole screened by the branes). We
compare our numerical calculations with the analytic solution.

In Section~\ref{sec:transition} we consider de Sitter (inflating)
branes which are initially in a static configuration.  It turns
out that we often observe an instability of the inflating branes.
Analytic calculations of small scalar perturbations around this
background geometry show that the radion mass square $m^2$ for this
case can be negative \cite{tachyon}. A strong tachyonic instability
predicted analytically is in full agreement with the instability of
inflating branes found numerically.  For certain configurations of
potentials $V(\phi), U_i(\phi)$, we find the existence of two warped
geometry solutions with different values of the 4d cosmological
constant $\Lambda_4$ (i.e.\ the curvatures of the 4d slices).  The
brane configuration with the higher 4d curvature is in general
unstable due to this tachyonic radion mode, and violently
re-configures to the second static configuration with lower 4d
curvature.  We illustrate this effect with numerical simulations as
well as analytic calculations, see also \cite{tachyon}.

In Section~\ref{sec:boom} we give an example where the instability of
the brane configuration causes a brane collision.  In
Subsection~\ref{sec:boom1} we show that the space-time metric of the
5d geometry between colliding branes becomes homogeneous, i.e.
$y$-independent.  The time-dependent solutions asymptotically cease to
feel the scalar field potentials $V(\phi)$ and $U_i(\phi)$, and
approach a universal asymptotic.  It sounds natural {\it a posteriori}
that this universal asymptotic is nothing but a Kasner-like asymptotic
with a scalar field, which we describe in Subsection~\ref{sec:kasner}.
The effect of the branes here is manifested by the fact that the
Kasner indices associated with the three brane directions are equal,
but different from that associated with the $y$ direction.  This is a
strong gravity regime, so it is not surprising that the 4d induced
metric on the brane is different from that derived with moduli
approximations in terms of 4d effective theory.

In the Conclusion we summarize the most interesting physical results.
 Technical details are collected in several Appendices.

\section{Setup}
\label{sec:setup}

The class of braneworld models we are interested in is characterized
by the following action
\begin{eqnarray}
\kappa_5^2 \,  S &=& \frac{1}{2} \int d^5 x \sqrt{-\,g} \, R + \int d^5 x
\sqrt{- g} \, \left[ - \, \frac{1}{2} \left( \partial \phi \right)^2 -
V \left( \phi \right) \right] + \nonumber\\
&-& \sum_{i=1,2} \int_{b_i} d^4 x \sqrt{-\,\gamma} \, \left\{ \left[ K
\right] + {U}_i \left( \phi \right) \right\} \,\,,
\label{eq:system}
\end{eqnarray}
where $\kappa_5^2=\frac{1}{M_5^3}$ is a 5d gravitational constant.  In
this convention $\phi$ is measured in units of $\kappa_5^{-1}$ and
physical potentials are multiplied by $\kappa_5^{-2}$.  The first term
describes gravity in the five dimensional bulk space.  We use the
``mostly positive'' metric signature.  The second term corresponds to
a (minimally coupled) bulk scalar field with the potential $V(\phi)$.
The last term corresponds to two $3+1$ dimensional branes, which
constitute the boundary of the five dimensional space. We allow for a
potential term ${U(\phi})$ for the scalar field at each of the two
branes. We denote by $\gamma$ the induced metric on the two branes,
and by $K$ their extrinsic curvature. Here and in the following,
$\left[ Q \right] \equiv Q \left( y_+ \right) - Q \left( y_- \right)$
denotes the jump of any quantity $Q$ across a brane ($\pm$ defined
respect to the normal of the brane).  We assume $S^1/{\mathbb
Z}_2$ symmetry
across each brane.

The algorithm we have written is implemented for generic bulk and
brane potentials.  In this paper we specify the potentials introduced
for the brane stabilization \cite{gw}.  We choose
\begin{eqnarray}
\label{eq:pot}
V(\phi) &=& \frac{1}{2} \, m^2 \phi^2 + \Lambda \,\,, \nonumber\\
{U}_i (\phi)&=& \frac{1}{2}M_i \left( \phi_i - \sigma_i \right)^2 + \lambda_i \,\,,
\end{eqnarray}
where $\phi_i$ is the value of $\phi$ on the $i-$th brane. A 5d
cosmological constant $\Lambda$ in the bulk and tensions $\lambda_i$
on the branes are included in the potentials.

The two branes are assumed to be parallel. We denote by $y$ the
coordinate transverse to them, and by ${\bf x}$ the three spatial
longitudinal coordinates. We assume isometry along three dimensional
${\bf x}$ slices including the branes.  We have to specify a metric
$g_{AB}$ that respects this symmetry.  In brane cosmology it is
customary to use the metric in the form $ds^2=-n^2 dt^2 + a^2 d{\bf
  x}^2 +b dy^2$, where the metric components $n, a, b$ depend on $(t,
y)$. However, this form of metric does not exhaust the freedom of the
coordinate choices. Most significantly, in this metric the branes do
not stay at the fixed positions; in general $y_i=y_i(t)$. There are
other gauge choices, which were used for specific braneworld problems,
for example coordinates comoving with one of the branes, the choice of
the bulk scalar $\phi$ as the $y$-hypersurface and others.  In these
contexts, a gauge in which the position of one of the two branes
is time dependent was often preferred, and identified with the radion
field ${\cal R} \left( t \right)$ associated with the extra dimension.
Although this choice may lead to an easier interpretation of the
interbrane distance, the resulting bulk and junction conditions (see
below) are significantly more complicated. In addition, in terms of
the four dynamical quantities $a, b, n, \phi,$ and ${\cal R}\,$ the
system is actually under-determined, and some gauge fixing is needed
to have a closed set of equations.


For numerical simulations, it is preferable to have coordinates where
neither brane is moving, although it is not obvious {\it a priori}
that such a gauge can be constructed without loss of generality. It is
possible to choose coordinates such that the bulk metric has the ``2d
conformal gauge'',
\begin{equation}
d s^2 ={\rm e}^{2\,B \left( t, y \right)} \left(- d t^2+ d y^2 \right)
 + {\rm e}^{2\,A \left( t, y \right)} \, d{\bf x}^2\,\,,
\label{line}
\end{equation}
This gauge still has the residual freedom to change $(t, y) \to (t',
y')$ in a way that preserves the 2d conformal form.  It can be
demonstrated that this freedom can be used  to fix the
position of the two branes along $y\,$.
Without loss of generality, we can
locate them at $y=0,1\,$.  We found that in the 2d conformal gauge the set
of bulk equations~(\ref{eq:eom}) acquires a relatively simple form,
which is well suited for the numerical scheme we have adopted (see the
next section).  The possibility of choosing a gauge, in which the
metric is 2d conformal and in which the branes are at a fixed position
is shown explicitly in Appendix~\ref{sec:comotion}.  As it is discussed
in Appendix~\ref{sec:residual_gauge}, even these requirements do not
fix the gauge choice completely.


Although in the system of coordinates we have chosen the branes are
always at a fixed position along the $y$ axis, their physical distance
is encoded in the metric component $B\,$, which is a time dependent
quantity. Clearly, the distance between two extended objects is not an
invariant quantity in general relativity, and different definitions
can be adopted when they are in relative motion. A simple heuristic
possibility, which we adopt here, is to integrate the line element
across the extra dimension at a fixed time
\begin{equation}
D \left( t \right) \equiv \int_0^1 dy \, \sqrt{g_{55}} = \int_0^1 dy
\, {\rm e }^{B \left( t ,\, y \right)}\ .
\label{d}
\end{equation}
One can check that $D \left( t \right)$ is invariant under the
residual gauge freedom in our   coordinates (\ref{line}),
 which is discussed in Appendix~\ref{sec:residual_gauge}
(but not under general coordinate transformations).

For the output of our numerical calculations, we rely on gauge
invariant quantities such as the invariants of the 5d Weyl tensor
$C_{ABCD}C^{ABCD}$, the curvature scalar $R$ and others.  These
invariants are calculated using the metric in the form (\ref{line}).
Additionally, we can use the $4+1$ split of the 5d curvature,
symbolically written as $R=R_4+K^2$, where $R_4$ is the curvature of
the 4d slices. This will be especially useful when we work with de
Sitter (inflating) branes of constant curvature.

In the gauge we have chosen, the nontrivial five-dimensional Einstein
equations can be split into three dynamical equations
\begin{eqnarray}
&&\ddot{A} - A'' + 3 \dot{A}^2 - 3 A'^2 = \frac{2}{3}\, {\rm e}^{2 B}
V \,\,, \nonumber\\
&&\ddot{B} -B'' - 3 \dot{A}^2 + 3 A'^2 = - \frac{\dot{\phi}^2}{2} +
\frac{\phi'^2}{2} - \frac{1}{3} {\rm e}^{2 B} V \,\,, \nonumber\\
&&\ddot{\phi} - \phi'' + 3 \dot{A} \dot{\phi} - 3 A' \phi' + {\rm
e}^{2 B} V_{,\phi} = 0 \,\,,
\label{eq:eom}
\end{eqnarray}
plus two constraint equations
\begin{eqnarray}
\label{eq:constraint}
&&- A' \dot{A} + B' \dot{A} + A' \dot{B} - \dot{A}' =
\frac{1}{3} \, \dot{\phi} \, \phi' \,\,, \nonumber\\
&&2 A'^2 - A' B' + A'' - \dot{A}^2 - \dot{A} \dot{B} =
- \frac{\dot{\phi}^2}{6} -
\frac{\phi'^2}{6} - \frac{1}{3} \, {\rm e}^{2 B} V \,\,.
\end{eqnarray}
Dots and primes denote derivatives with respect to $t$ and $y\,$,
respectively. It is easy to show that the constraint equations are
preserved by the dynamical equations.

In addition, from the boundary terms in the action for the two branes
we recover the following junction (Israel) conditions
\begin{eqnarray}
\label{eq:bc}
\left[ A' \right] &=& \mp \frac{1}{3} \, {U} \, {\rm e}^B\ ,
\nonumber\\
\left[ B' \right] &=& \mp \frac{1}{3} \, {U} \, {\rm e}^B\ ,\nonumber\\
\left[ \phi' \right] &=& \pm {\rm e}^B \, U_{,\phi}
\,\,,
\end{eqnarray}
where the upper and lower signs refer to the branes at $y=0,1\,$,
respectively. We impose $\mathbb{Z}_2$ symmetry across the two branes.
That is, for any given function $Q$, we assume that
\begin{equation}
\left[ Q^\prime \right]_0 = 2 \, Q^\prime \left( 0^+ \right)
\quad,\quad\quad \left[ Q^\prime \right]_1 = - \, 2 \, Q^\prime \left(
1^- \right) \,\,.
\end{equation}

To conclude, we describe the four dimensional induced metrics of the
two branes. Since they are at fixed positions, their induced metrics
are simply given by
\begin{equation}
d s^2 = - d \tau^2 + a^2 \left( \tau \right) d {\bf x}^2 \,\,.
\label{ind}
\end{equation}
That is, we recover a FRW Universe with proper time $d \tau = {\rm
  e}^{B_i} d \, t\,$, and scale factor $a = {\rm e}^{A_i}\,$ (where
$A_i$ and $B_i$ refer to quantities evaluated at the positions of the
two branes). The Hubble parameters on the two branes are thus given by
\begin{equation}
H_i \equiv \frac{1}{a} \, \frac{d\,a}{d\,\tau} \Big\vert_i = {\rm
e}^{-B_i} \, \dot{A}_i \,\,,
\label{hubble}
\end{equation}
where, as usual, dot denotes derivative with respect to the bulk time
$t\,$. The Hubble parameters $H_i$ are invariant under residual gauge
transformations of the metric (\ref{line}).

\section{Numerical Code}
\label{sec:numerics}

In this section, we describe the algorithm that we employ to integrate
the equations of motion \eqref{eq:eom} numerically. The algorithm
copes with two tasks: It provides the time evolution of $N+1$ grid
sites, equally spaced between the two branes at $y=0,1\,$, and it
solves the constraints arising from the boundary conditions at the two
branes. In both cases, a second order discretization scheme is used.
In the current version of the program, the same step size of
discretization is employed in both the time and spatial directions:
$dt = dy = 1/N \equiv \epsilon\,$. This assumption is made not only
for simplicity, but also to assure the proper propagation of the
numerical data along the characteristics of the partial differential
equations.

It is convenient to scale the factor $\frac{1}{\kappa_5^2}=M_5^3$ out
of the action (\ref{eq:system}). This fixes the units of the scalar
field $\phi$ and its potentials. However, the dynamical equations of
motion admit a scaling of the metric functions, which allows to
choose, in principle, arbitrary units of the space-time scales. We use
this freedom to secure the supergravity limit of our models, this is
to say that all length scales are greater than the 5d fundamental
scales $l_5=M_5^{-1} $.  We discuss this issue at a greater length
in Section \ref{sec:units}.

\subsection{Bulk Evolution}
\label{sec:numerics.bulkevol}

The system of bulk equations consists of the three second order
differential equations~(\ref{eq:eom}) for the functions $A$, $B$, and
$\phi\,$, which we call bulk evolution equations, as well as the two
constraint equations~(\ref{eq:constraint}). The latter are preserved
by the evolution equations, and can be used as a check of accuracy of
the numerical integration.

In the evolution equations \eqref{eq:eom}, derivatives of functions
only appear in the forms $\ddot{f} - f^{\prime\prime}$ and ${\dot f}
\, {\dot g} - {f^\prime} {g^\prime}\,$.

\begin{figure}[h]
  \centering
  \epsfig{file=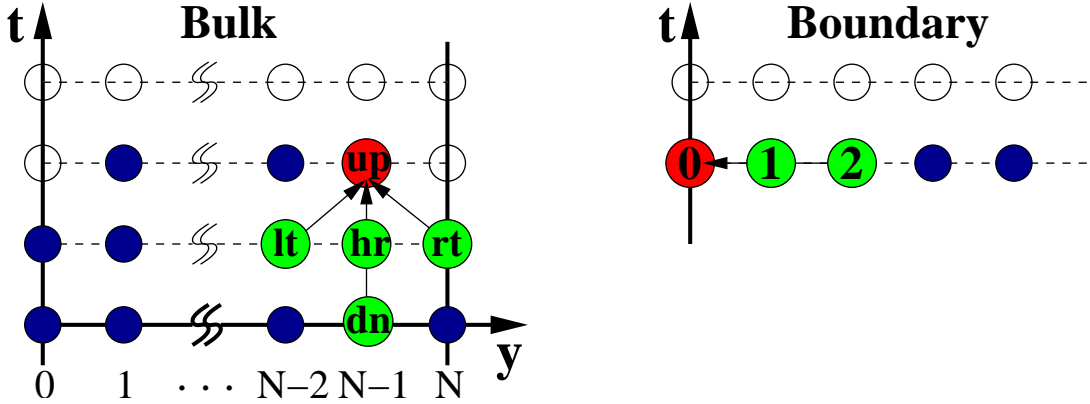,width=14.5cm}
  \caption{Numerical evolution scheme.}
  \label{fig:num_evolution}
\end{figure}

We discretize the equations by finite-differencing these combinations
using the leapfrog scheme (see Figure~\ref{fig:num_evolution}). Let
$f_{\text{hr}}$ denote the value of the field $f$ at a given grid
point $y_i\,$ on the last time step $t$ that had been computed,
$f_{\text{hr}} \equiv f \left( t, y_i \right)\,$. Then define
$f_{\text{lt}}$ and $f_{\text{rt}}$ to be the value of $f$ at the same
time $t$ and on the left and right neighboring sites, $f_{\text{lt}}
\equiv f \left( t, y_{i-1} \right)$ and $f_{\text{rt}} \equiv f \left(
  t, y_{i+1} \right)\,$. Finally, denote with $f_{\text{dn}}$ and
$f_{\text{up}}$ the value of the function on $y_i$ at the two times
just before and after $t\,$, respectively $f_{\text{dn}} \equiv f
\left( t - d t, y_i \right)$ and $f_{\text{up}} \equiv f \left( t +
  dt, y_i \right)\,$ (see Figure~\ref{fig:num_evolution}). In terms of
these quantities, the relevant differential operators of $f$ at the
point $\left( t, y_i \right)$ can be discretized with second order
accuracy as
\begin{eqnarray}
\ddot{f}  - f^{\prime\prime} &=&  \frac{1}{\epsilon} \lmk  f_\te{up} +
f_\te{dn} - f_\te{lt} - f_\te{rt}  \rmk + {\rm O} \lmk \epsilon^2 \rmk
\,\, , \nonumber\\
{\dot f} \, {\dot g} - {f^\prime} {g^\prime} &=& \frac{1}{4 \,
\epsilon^2} \lkk \lmk f_\te{up} - f_\te{dn} \rmk \, \lmk g_\te{up} -
g_\te{dn} \rmk - \lmk f_\te{rt} - f_\te{lt} \rmk \, \lmk g_\te{rt} -
g_\te{lt} \rmk \rkk + {\rm O} \lmk \epsilon^2 \rmk \,\,.
\label{discrete}
\end{eqnarray}
Recall that $\epsilon = 1/N$ corresponds to the distance
between consecutive grid sites. After the discretization, the three
evolution equations become three algebraic equations, which can be
solved for the unknown quantities $A_\te{up}$, $B_\te{up}$, and
$\phi_\te{up}$. This procedure is repeated at each bulk site, leading to
the bulk values of the three functions at $t + d t\,$, which are then
used in the subsequent time steps.
 
\subsection{Boundary Conditions}
\label{sec:numerics.bc}

The numerical scheme described in the previous subsection allows us to
determine the value of the metric coefficients and of the scalar field
at the next time-step for all the bulk sites, but not for the two
sites $i=0,N\,$, corresponding to the positions of the two branes. To
obtain the latter, the boundary conditions~\eqref{eq:bc} have to be
used. First we advance all the bulk sites as described in the previous
section. Once we know the value of $A$, $B$, and $\phi$ in the bulk at
time $t+dt$, equations~\eqref{eq:bc} can be finite-differenced into a
set of algebraic equations for the boundary values at that time. In
the following we describe how to implement this procedure at $y=0\,$.
The computation for the other brane proceeds analogously.

The boundary conditions contain first derivatives with respect to $y$
of the metric coefficients and of the scalar field at the brane
locations. An asymmetric discretization for the first derivative of a
generic function $f\,$, which preserves second order accuracy in
$\epsilon\,$, is given by
\begin{eqnarray}
f^\prime_0 &=& \frac{1}{2 \, \epsilon} \lmk - \, 3 \, f_0 + 4 f_1 -
f_2 \rmk + {\rm O} \lmk \epsilon^2 \rmk \,\,.
\label{eq:bc_difference}
\end{eqnarray}

Since the right hand side of the first two boundary conditions
coincide, we replace the first of them simply by $A^\prime_0 -
B^\prime_0 = 0 \,$, or, using equation~(\ref{eq:bc_difference}),
\begin{equation}
- 3 \lmk A_0 - B_0 \rmk + 4 \lmk A_1-B_1 \rmk- \lmk A_2 - B_2 \rmk = 0
  \,\,.
\label{eq:bc_diff_a}
\end{equation}
If we define $\beta \equiv e^B\,$, the second boundary condition
simplifies to $(1/\beta)^\prime= {U}_0 \left( \phi_0 \right) /
6\,$, which can be rewritten as
\begin{equation}
\beta_0 = \frac{3 \, \beta_1 \, \beta_2}{4 \, \beta_2 -\beta_1 -
\frac{\epsilon}{3} \, {U}_0 \left( \phi_0 \right) \, \beta_1 \,
\beta_2} \,\,.
\label{eq:bc_diff_b}
\end{equation}
Finally, the third boundary condition gives
\begin{equation}
4 \, \phi_1 - \phi_2 - 3 \, \phi_0 - \epsilon \, \beta_0 \, {U}_0^\prime(\phi_0) = 0 \,\,.
\label{eq:bc_diff_c}
\end{equation}

Only the values of the three functions on the brane are unknown.  By
substituting the value for $\beta_0$ given by
equation~(\ref{eq:bc_diff_b}) into equation~(\ref{eq:bc_diff_c}), the
latter becomes an equation where the only unknown quantity is
$\phi_0\,$. For specific brane potentials ${U}_0\,$, this equation can
be solved analytically; more generally, one can solve it numerically
through some iterative method. In our algorithm, the iterative
Newton's method is employed. Finally the value of $B_0 = \ln(\beta_0)$
can be used in \eqref{eq:bc_diff_a} to determine $A_0\,$.

\subsection{Initial Configurations}
\label{sec:static}

Initial conditions are imposed by specifying the three functions and
their first time derivatives on the grid sites at some initial time $t_0
\equiv 0\,$.  We denote them as $A_0 \left( y_i \right)$, $\dots$,
$\dot{\phi}_0 \left( y_i \right) \,$, with $i$ ranging from $1$ to
$N-1\,$  in the bulk ($i=0,N$ are the sites of the two branes). These functions can
 not be chosen arbitrarily but rather must satisfy the constraint equations
 (\ref{eq:constraint}).  Once this is done a second order
Runge-Kutta time step is used to ``convert'' the initial conditions of
the form $f_0 \left( y \right)$, $\dot f_0 \left( y \right)$, into
initial conditions given at the first two initial time steps $f_0
\left( y \right)$ and $f_{0+dt} \left( y \right)$. The Runge-Kutta
step is done as follows
\begin{equation}
\label{eq:rk2}
f_{0+dt} \left( y \right) = f_0 \left( y \right) + dt \left( \dot f_0
\left( y \right) + \frac{1}{2} dt \, \ddot f_0 \left( y \right)\right)
\,\,,
\end{equation}
where the second time derivatives $\ddot f_0(y)$ are replaced by the
equations of motion \eqref{eq:eom}. This ``conversion'' is needed to
have the initial conditions in a form suitable for the leapfrog scheme
described in Subsection~\ref{sec:numerics.bulkevol}.

In general, the initial time derivatives can be non-vanishing, so that
one can study situations in which the geometry of the extra dimension
is time-dependent already at the beginning of the numerical
integration. 
For example, 
this is the case for
 the AdS--Schwarzschild solution we will deal with 
in Section~\ref{sec:ads-schw}.
 For each such case  the  choice
 of initial conditions must be 
consistent with the constraint equations (\ref{eq:constraint}).
 We discuss one such algorithm  in Section~\ref{sec:ads-schw}.

A particularly interesting class of
initial conditions is however the one of static warped
solutions
\begin{equation}\label{warp}
  ds^2 =  W(y)^2 \left( dy^2  - dt^2 + e^{2Ht}d{\bf x}^2 \right).
\end{equation}
characterized by a fixed bulk geometry and maximally symmetric
(de Sitter or Minkowski) branes. 
This metric turns to the
form~(\ref{line}) with the identification
\begin{equation}
B \left( t, y \right) \rightarrow B \left( y \right)=\ln W \ ,
\quad\quad\quad
A \left( t, y \right) \rightarrow B \left( y \right) + H \, t \ ,
\label{station}
\end{equation}
where $H$ is the Hubble parameter of the de Sitter brane and the bulk
scalar field $\phi$ is also a function of $y$ only. Such solutions
were studied with dynamical system methods in~\cite{FFK}. The
numerical integration can be used to check their stability. Numerical
errors due to the grid discretization act as small perturbations. If
the initial configuration is not stable, the tiny numerical errors
accumulate with time, and eventually lead to an evolution of the
system. When this happens, a full numerical calculation is the only
tool to study where this evolution leads to, namely whether the two
branes collide, move apart to infinity, or get stabilized at a finite
distance in another static but stable configuration.  As we
will see below, in many cases de Sitter branes turn out to be
unstable. Therefore even static warped geometry configurations can
provide suitable initial conditions for time-dependent braneworld
dynamics!

 When numerical inaccuracy is used to seed the
 evolution, as described above, the  initial  amplitude and consequently the timing
 of the  instability depends on the
accuracy of the numerical integrator. This accuracy is in turn related to the
spacing of the grid sites in the bulk. Increasing the number of grid
sites decreases their separation, and the instability develops later.
Alternatively, initial perturbations on the top of the static
configurations can be imposed directly as initial conditions. This
allows a quicker development of the instability, or, for static
configurations, the excitation of some of the lowest  eigenmodes
of the system. A simple class of initial perturbations, which can be
implemented in our numerical algorithm, is described in
Appendix~\ref{sec:Perturbations} and illustrated in
Figure~\ref{fig:instability}.  We found, however,
 that the qualitative behavior of the system did not depend
 on the details of how the initial perturbations are generated, whether
 imposed explicitly or through numerical roundoff errors.
 In the following, we therefore discuss instead how
the static configurations are determined.

For static configurations, the bulk equations reduce to
\begin{eqnarray}
\label{eq:static}
&&\phi^{\prime\prime} + 3 \, B^\prime \, \phi^\prime - {\rm e}^{2 B}
V^\prime \left( \phi \right) = 0 \,\,, \nonumber\\
&& {B^\prime}^2 + \frac{1}{6} \, {\rm e}^{2 B} \, V \left(\phi \right)
- \frac{1}{12} \, {\phi^\prime}^2 = H^2 \,\,.
\end{eqnarray}
In addition, the last two of the boundary conditions~(\ref{eq:bc}) have to be satisfied at
each brane. In the gauge we are using here the constraint equations
 (\ref{eq:constraint}) are automatically satisfied.
 
 The bulk equations are thus reduced to a system of first order
 differential equations for the functions $B\,,\phi\,$, and $\phi'\,$,
 so that the phase space of possible solutions is effectively
 three-dimensional~\cite{FFK}. To solve these equations subject to
 (given) boundary conditions, we specify the values of the three
 functions at $y=0\,$, as well as the value of the constant parameter
 $H\,$\footnote{One may be wondering why we can specify four variables
   in a 3D phase space. Recall, however, that in our gauge the
   position of the second brane is fixed at $y=1$. In the language
   of~\cite{FFK} we are using three degrees of freedom to specify the
   starting point of our trajectory in phase space and one to specify
   the length of the trajectory, i.e.\ at what point on the trajectory
   the second brane will be found.}. For a given brane potential
 $U_0\,$, only two of these four numbers can be chosen arbitrarily,
 and the other two are determined by the junction conditions at the
 first brane. (In the 3d phase space this means that the junction
 condition at the first brane defines a 1d curve in phase space along
 which the trajectory must begin.) The bulk
 equations~(\ref{eq:static}) are then integrated with a standard
 fourth order Runge-Kutta integrator. Depending on the initial values
 and on the bulk potential, the bulk solution may become singular
 before the brane at $y =1$ is encountered. If this happens, some
 other initial values (or some other bulk potential) have to be
 considered.

Even if the brane at $y=1\,$ is reached, we face the nontrivial
problem of satisfying the boundary conditions also at the second
brane. The simplest way to solve it is to regard the junction
conditions as equations for the parameters of the brane
potentials. One can freely choose the three numerical values at the
first brane (as well as the numerical value of $H$), integrate the
bulk equations, and then use the junction conditions to determine the
potentials at the two branes.~\footnote{In general, this does not
determine the brane potentials, but only their values and their
derivatives at a single value of the field $\phi\,$. One can complete
the functional form of the potential arbitrarily, say as in
equation~(\ref{eq:pot}). In this case, one is for example free to choose large
positive values for the two mass parameters $M_i\,$, favoring
values of the scalar field at the branes which are close to the vacuum
expectation values $\sigma_i\,$.}  However, one is typically interested
in the more difficult situations in which the brane potentials are
specified, and the initial configurations have to be determined
accordingly.

In the second case, we face a boundary-value problem:
values of the fields satisfying the boundary conditions at the first brane do not in
general lead to field values that satisfy the boundary conditions at the second brane, once
they are evolved across the bulk according the bulk differential
equations~(\ref{eq:static}). It is by no means guaranteed that
any choices consistent with the junction conditions at both branes exist.
 Indeed, as discussed in~\cite{FFK},
many potentials do not give static solutions at all, while some
others typically lead to only a finite number of them. In
Appendix~\ref{sec:shoot} we discuss the numerical method (known as the
``shooting'' method \cite{NR}) which we employ to find these solutions.

\subsection{Units}
\label{sec:units}

Let us inspect the dynamical equations (\ref{eq:eom}), constraint equations
(\ref{eq:constraint}), and the boundary conditions 
(\ref{eq:bc}). While the  units of the bulk scalar field are fixed by
 our form of the action (\ref{eq:system}), it is easy to see that
these equations are invariant under the following scaling transformation
\begin{equation}\label{scaling1}
 A \to A+S' \, , \,\,B \to B+S \, ,
\end{equation}
where $S'$ and $S$ are  arbitrary real valued transformation parameters.
The scalar field potentials enter the equations only in the combinations
$e^{2B} \, V$ and $e^B \, U$. Therefore  (\ref{scaling1})
can be accompanied by the transformation
\begin{equation}\label{scaling2}
V \to e^{-2S} \, V , \,\,\, U \to e^{-S} \, U .
\end{equation}

Suppose some metric functions $A, B$ are the solutions of (\ref{eq:eom})
for given potentials $V, U$.  The scaling transformations (\ref{scaling1}) and (\ref{scaling2})
tell us that from these metric functions we can generate a family of solutions
for re-scaled potentials. 
This is very useful for introducing the units of  scales for numerics. 
Indeed, while $y$ and $t$ in (\ref{line}) are 2d
conformal length and time (i.e. affine parameters along corresponding 
directions),
the metric function $e^B$ defines the physical
interbrane distance $D$ and the physical time. 
As often occurs in numerical simulations, it is not always easy to 
extend the range of variables, like $e^B$ in our case. 
As we will see in the example of the next Section, numerical stability
(without brane stabilization) has a controlled but finite life time.
If we naively increase the scale of $e^B$, the stability will be
short-lived. 
The trick is to continue to work with numerically convenient 
values of $e^B$, but interpret scales in units of $l=e^S \, l_5$.
One can take $S$ to be large enough to have the scale $e^S \, l_5$
much greater than the fundamental bulk scale $l_5$. 
This is to say that numerically we solve our equations not only for a given
scale and given choice of parameters of the potential,
but for the whole family of scales and parameters
which corresponds to the orbit of the group transformation 
(\ref{scaling1}), (\ref{scaling2}).
For the parameters of the potentials
(\ref{eq:pot}) we have the following units:
$[m]=e^{-S}M_5$, $[M]=e^{-S}M_5$, $[\Lambda]=e^{-2S}M_5^5$,
$[\lambda]=e^{-S}M_5^4$, $[\phi]=[\sigma]=M_5^{-3/2}$.

The time evolution of variables in the paper will be plotted
versus conformal time $t$. The units of $t$ are the light crossing
time between branes. This corresponds to the distance between branes in
the conformal coordinate $y$ which is  simply $1$ in our units.

As usual, the parameters for the numerical simulations do not allow
the introduction of a large hierarchy, since numerical inaccuracies
accumulate much faster. Therefore most of the parameters are chosen to
be of order unity in our units.  The values of parameters that
correspond to the numerical runs shown in the figures of this paper are
collected in the Table \ref{tab:parameters} in the Appendix.

\section{Branes in AdS Background Without a Scalar Field}
\label{sec:rs}

The algorithm presented above allows an exact integration of general
two-brane configurations with bulk scalar fields. In this and in the
next sections we discuss in detail several applications. The first two
examples of this section have no scalar field, and the evolution is
known analytically. We report them mainly to discuss the accuracy of
the code and to outline its main features.  The code is accurate
enough to reproduce the known analytic solutions.  In the following
sections we will study the more complicated non-linear evolution of a
system with a scalar field, for which the solutions were not
previously known.  Fortunately, we still will be able to check certain
properties of the solutions analytically.

In this section we first consider the static unstabilized
Randall-Sundrum flat brane solution, from the point of view of the
numerical solution of the equations.  Then we study non-static
(moving) branes in an AdS-Schwarzschild background, and compare the
numerical solution with the known AdS-Schwarzschild solution.

\subsection{The Randall--Sundrum Model}
\label{sec:rs1}

Our first example is the two-brane Randall-Sundrum model \cite{RS1}.
It represents a particularly simple example of a brane world that only
consists of a five dimensional AdS space with a curvature radius
$l^2=-6/\Lambda$, determined by its 5d cosmological constant
$\Lambda$, and of two flat branes with tensions $\lambda_i=\pm 6/l$ .
The system is entirely described by one time independent function. In
terms of the 2d conformal gauge (\ref{line}) we have
\begin{equation}
\label{eq:rs_sol}
A \left( y \right) = B \left( y \right) = - \ln \left[ \frac{y +
\left({\rm e}^{D/l} - 1 \right)^{-1}}{l} \right] \,\,,
\end{equation}
where $D\,$ can take any constant value, which - according to
equation~(\ref{d}) - corresponds to the interbrane distance.
\begin{figure}[h]
\begin{minipage}{16.5cm}
\begin{minipage}[b]{8.25cm}
\begin{flushleft}
  \epsfig{file=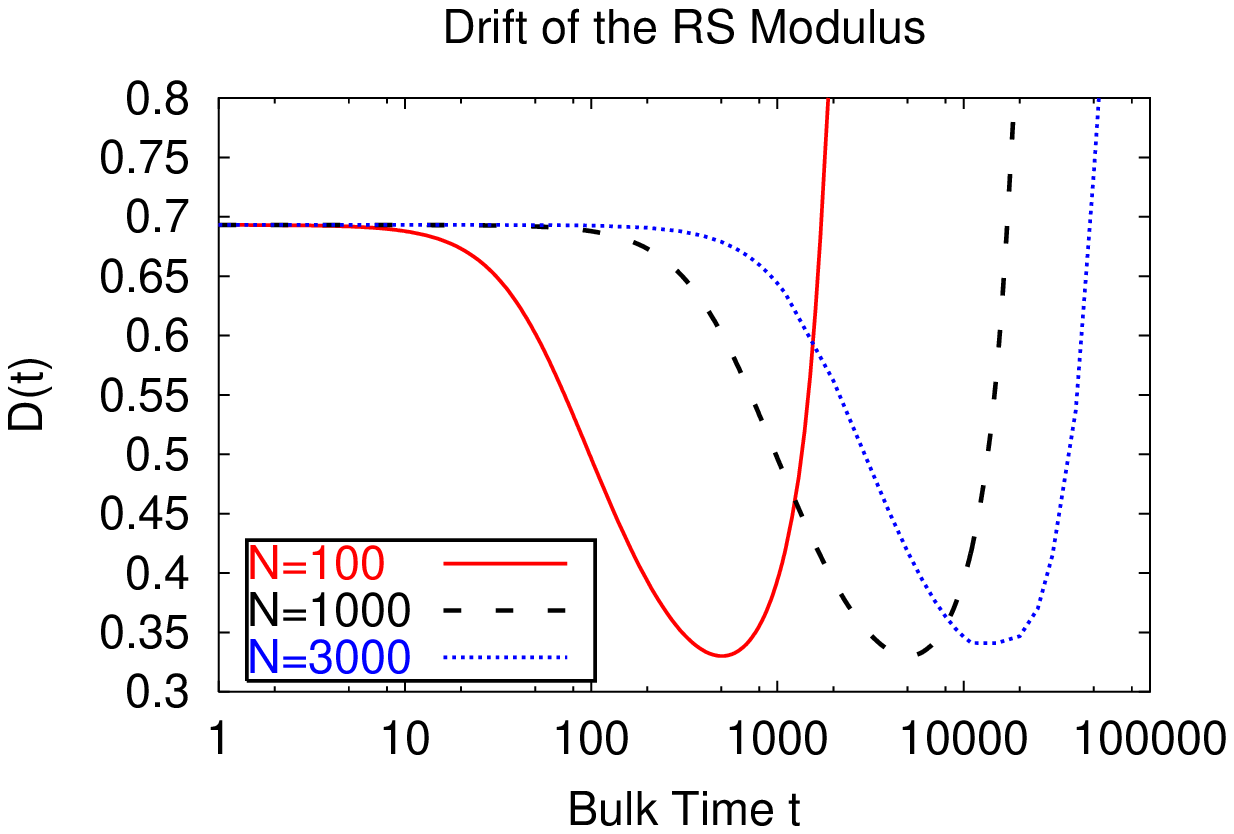,width=8.25cm}\\
\end{flushleft}
\par
\end{minipage}\hfill
\begin{minipage}[b]{8.25cm}
\begin{flushright}
   \epsfig{file=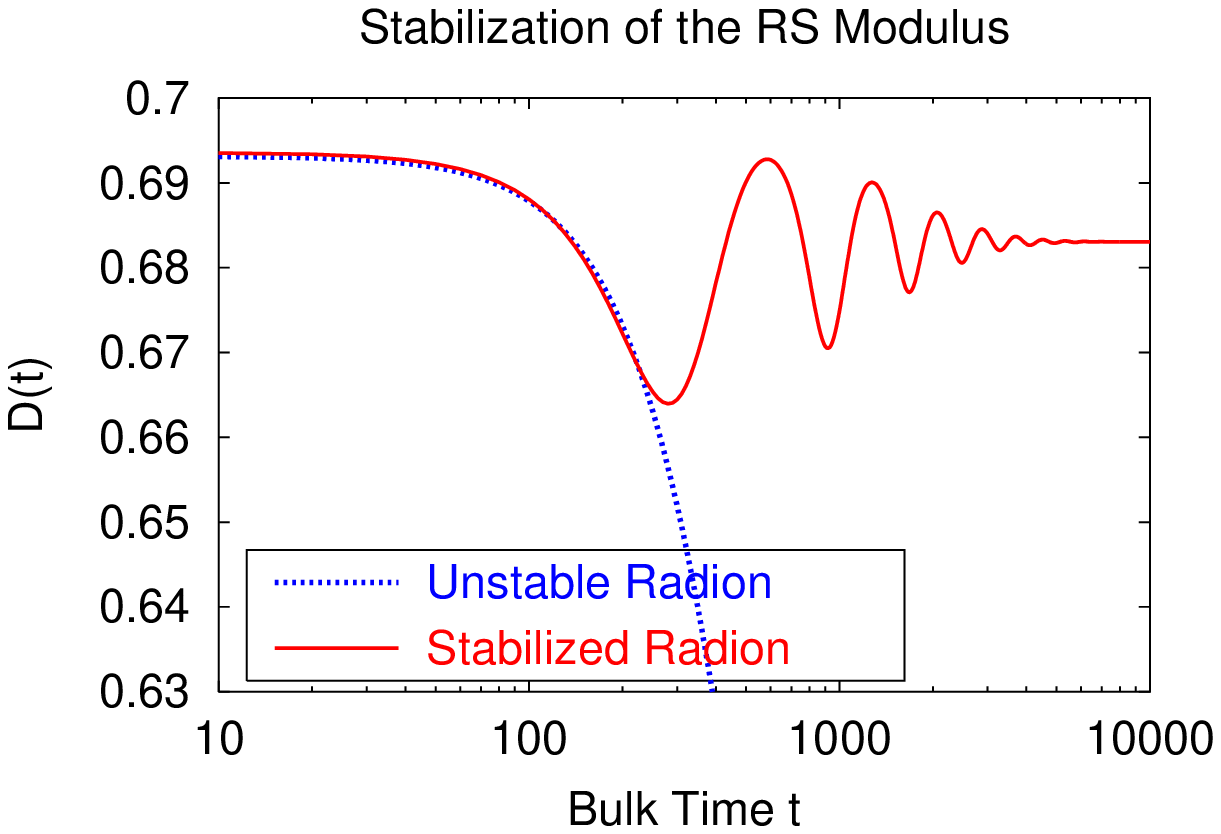,width=8.25cm}\\
\end{flushright}
\par
\end{minipage}
\end{minipage}
\caption{Left Panel: Drift of the free modulus of the RS model depending on the
  numerical accuracy. Right Panel: Stabilization of the RS radion.}
\label{fig:rs}
\end{figure}
We can reproduce this set-up in our code by simply setting to zero all
the scalar field related parameters in the bulk/brane potentials in
equations (\ref{eq:eom}) (as well as the initial conditions for the
scalar field).  The numerical solutions of the equations
(\ref{eq:eom}) are in agreement with (\ref{eq:rs_sol}). 
%
%
As discussed in section~\ref{sec:static}, small perturbations are
unavoidably introduced by the discretization. Thus, this set-up is
particularly useful for verifying the accuracy of the code.  Notice
also, that numerical instability is much worse for the RS without
stabilization.  In the left panel of Figure~\ref{fig:rs} we show how
the time-scale at which the instability develops is related to the
number $N$ of bulk sites. The more we increase $N$ the more the
accuracy of the computation increases, and numerical instability is
delayed for the later times.  We estimate the time scale where the
code is stable as being proportional to the grid resolution $N$.

The right panel of Figure~\ref{fig:rs} also shows how the introduction
of the stabilization mechanism (with the bulk scalar fields with the
potential~\eqref{eq:pot}) can, for appropriate choices of the
parameters, lead to a stabilization of the interbrane distance.  In
this his case the code is much more stable.  We discuss this issue in
more detail in Section~\ref{sec:transition}.

The choice of parameters and initial conditions that was used in the
numerical runs plotted in Figure~\ref{fig:rs} as well as the ones for
all following simulations are collected in Table~\ref{tab:parameters}.

\subsection{The AdS--Schwarzschild Solution}
\label{sec:ads-schw}

Starting from a setting similar to the Randall-Sundrum example of the
previous section, but allowing for non-vanishing initial (at $t=t_0$)
time derivatives, we generate time-dependent numerical solutions that
belong to a larger class of solutions. Assuming the initial spatial
profile $A \left( t_0, y \right) = B \left( t_0, y \right)$ of
equation~\eqref{eq:rs_sol}, the constraint
equations~\eqref{eq:constraint} are solved by
\begin{equation}
\label{eq:ads_constraint}
\dot A (t_0) = c \, \left[ y + \frac{1}{{\rm e}^{D/l} -1} \right]
\quad,\quad\quad \dot B(t_0) = - \dot A (t_0)  \,\, ,
\end{equation}
where $c$ is a constant.  The choice $c=0$ gives Randall-Sundrum
solutions, while a non-vanishing $c$ corresponds to moving branes.
From the Birkhoff theorem for plane-parallel brane configurations it
follows \cite{birk} that the generic 5d bulk metric must be a stripe
of the AdS--Schwarzschild geometry (where the Schwarzschild mass is
virtual because it is screened by the branes). Thus, the branes are
moving in an AdS--Schwarzschild background.
To see this, note that in the absence of the scalar field and for
brane tensions as in the Randall-Sundrum model, the boundary
conditions~(\ref{eq:bc}) give $A^\prime \left( A^\prime + B^\prime
\right) = 2 \, \Lambda \, {\rm exp} \left( 2 \, B \right)\,$ at the
location of the two branes. From the bulk equations, we then recover
\begin{equation}
\ddot{A} + 2 \, \dot{A}^2 - \dot{A} \, \dot{B} = 0 \quad,\quad\quad y
= 0 ,\, 1 \,\,,
\end{equation}
which, in terms on the proper time $\tau$ and the Hubble parameter $H_i$
on the two branes (cf.\ equations~(\ref{ind}) and~(\ref{hubble})), becomes
\begin{equation}
\frac{d \, H_i}{d \tau} + 2 \, H_i^2 = 0 \,\,.
\end{equation}
This corresponds to a radiation dominated standard four dimensional
Universe,
\begin{equation}
\label{eq:dark_rad_sol}
H = \frac{H_0}{1 + 2 \, H_0 \, \left( \tau-\tau_0 \right)} \,\,.
\end{equation}
%
%
The appearance of effective radiation domination on the branes is
characteristic of an AdS--Schwarzschild bulk geometry~\cite{adssch}.
The invariant of the 5d Weyl tensor $C^2=C_{ABCD}C^{ABCD}$ projected
into the brane scales as $C^2 \propto a^{-8}\,$, where $a \left( \tau
\right)\,$ is the scale factor of the induced FRW brane metric.  Since
$a \left( \tau \right)\,$ is radiation dominated, at the brane we have
\begin{equation}
\label{eq:weyl_sol}
C^2 = C_0^2 \left[ 1 + 2 \, H_0 \left( \tau - \tau_0 \right)
\right]^{-4} \,\,.
\end{equation} 

\begin{center}
\begin{figure}[h]
\begin{minipage}{16.5cm}
\begin{minipage}[b]{8.25cm}
\begin{flushleft}
   \epsfig{file=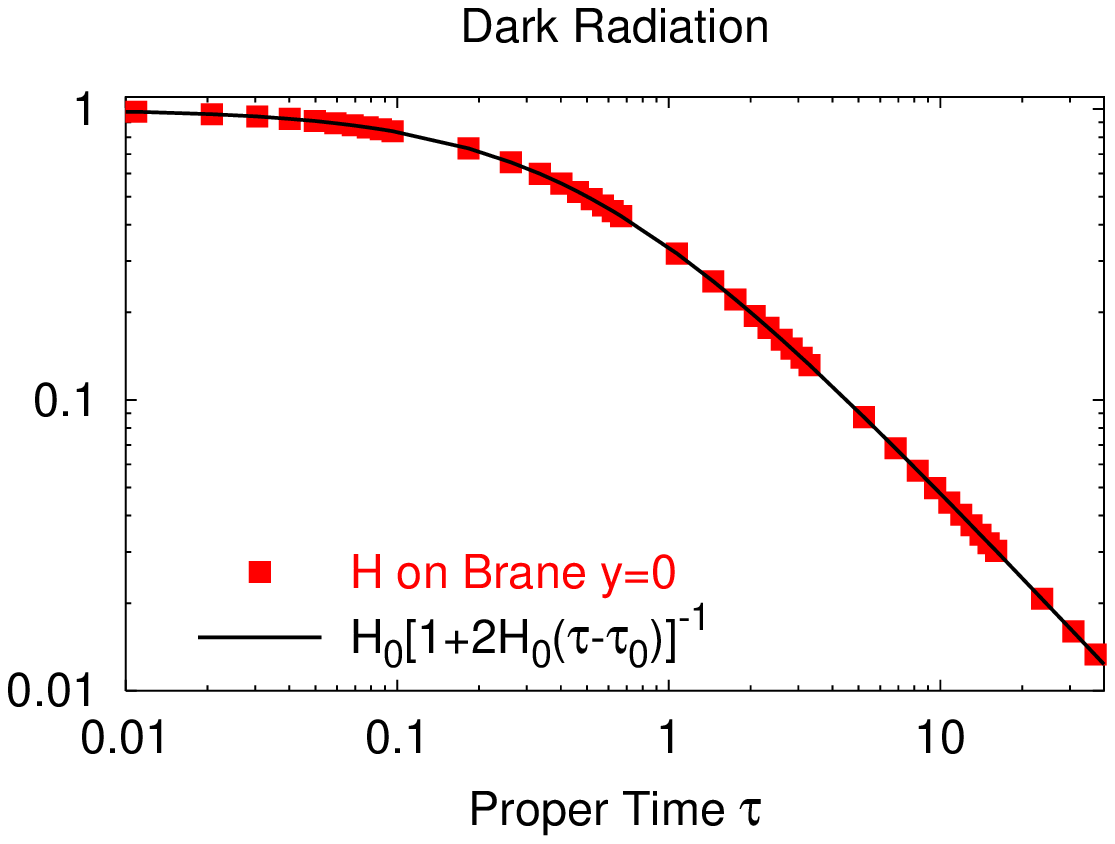,width=8.25cm}\\
\end{flushleft}
\par
\end{minipage}\hfill
\begin{minipage}[b]{8.25cm}
\begin{flushright}
  \epsfig{file=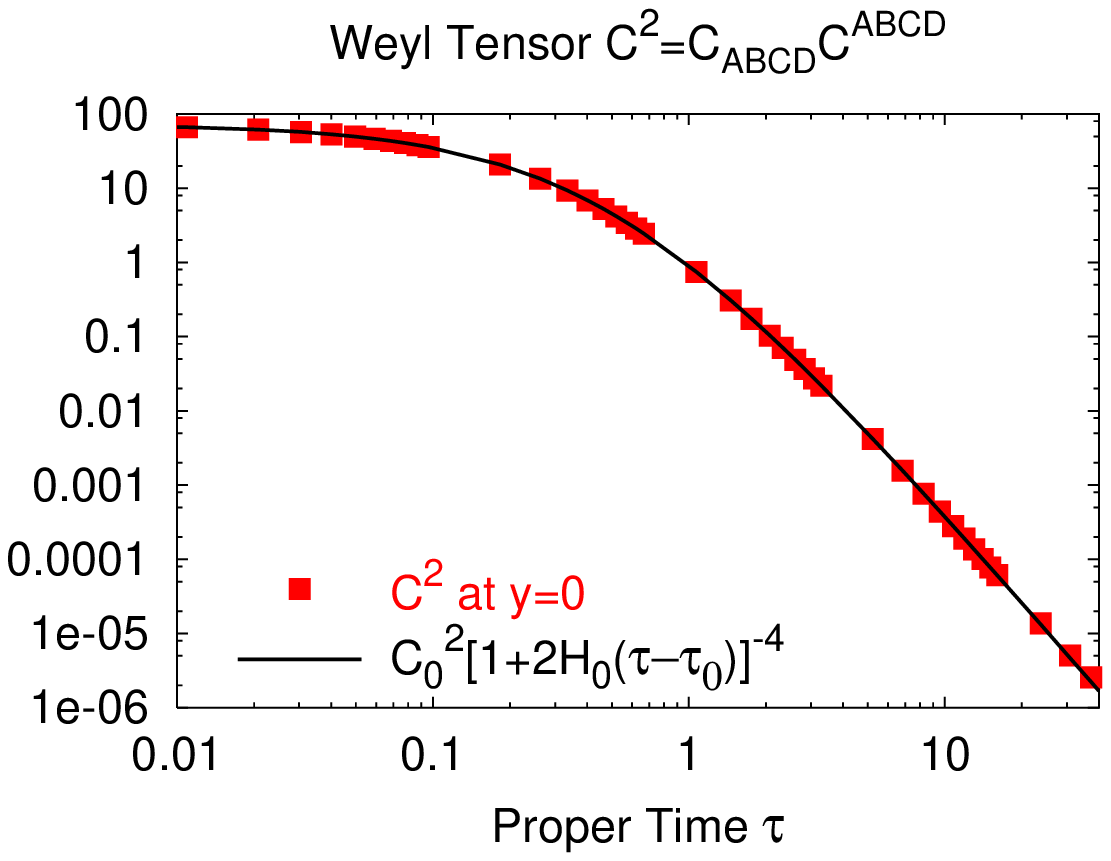,width=8.25cm}\\
\end{flushright}
\par
\end{minipage}
\end{minipage}
\caption{Comparison between
numeric and analytic solutions for 
 the Hubble parameter and Weyl tensor on
the brane.}
\label{fig:dark_radiation}
\end{figure}
\end{center}

The two equations~(\ref{eq:dark_rad_sol}) and~(\ref{eq:weyl_sol}) are
independent of the choice of coordinates in the bulk, and can be
easily reproduced with our code. For a particular realization of the
initial conditions (\ref{eq:ads_constraint}) with the parameter $c=1$,
in Figure~\ref{fig:dark_radiation} we plot the numerical calculation
(squares) of the time evolution of $H$ and the 5d Weyl tensor on the
brane.  Solid curves correspond to the AdS-Schwarzschild analytic
solutions (\ref{eq:dark_rad_sol}) and~(\ref{eq:weyl_sol}).  The
agreement between numerics and analytics is manifest.
\begin{center}
\begin{figure}[h]
\begin{minipage}{16.5cm}
\begin{minipage}[b]{8cm}
\begin{flushleft}
   \includegraphics[width=8.25cm,viewport= 25 15 320 205,clip]{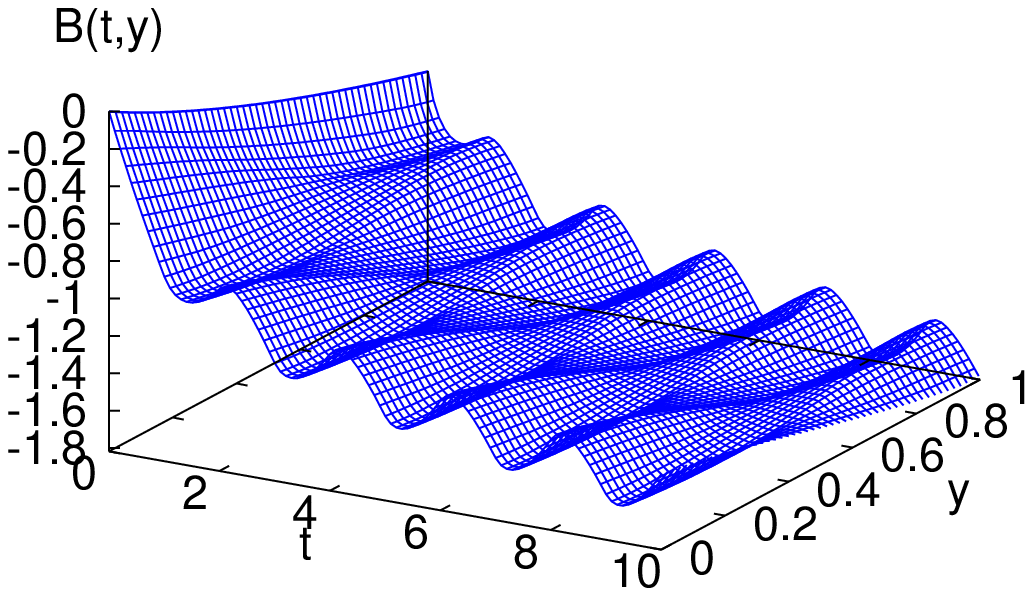}\\
\end{flushleft}
\par
\end{minipage}\hfill
\begin{minipage}[b]{8cm}
\begin{center}
  \epsfig{file=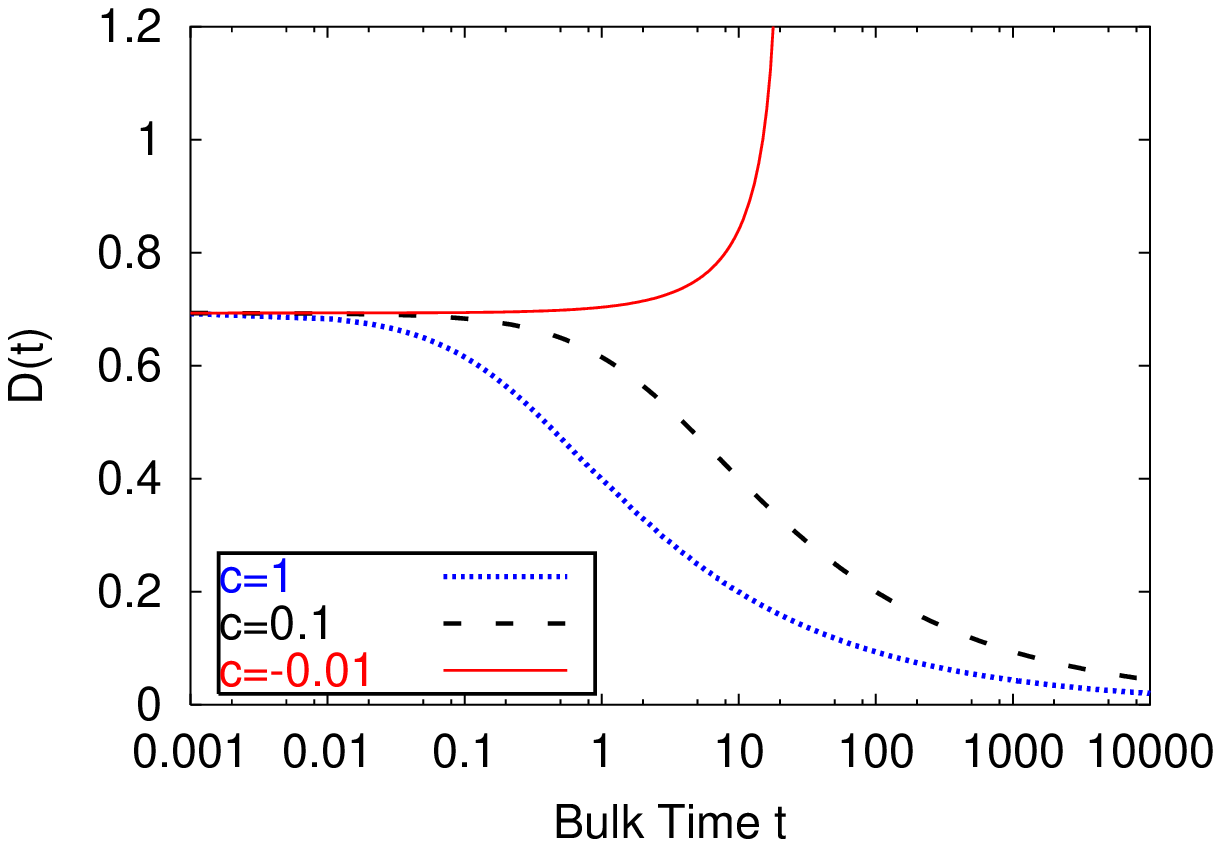,width=8.25cm}\\
\end{center}
\par
\end{minipage}
\end{minipage}
\caption{Left panel: Non-physical waves appearing as gauge modes in the metric
function $B \left( t ,\, y \right)$ (for $c=1$). Right panel:
 interbrane distance for
various initial conditions controlled by the parameter  $c$.}
\label{fig:ads_radion}
\end{figure}
\end{center}
In the left panel of Figure~\ref{fig:ads_radion}, we show instead the
evolution of the metric component $B \left( t,\,y \right)$ for the
same configuration used to generate Figure~\ref{fig:dark_radiation}.
The ripples of $B(t,\,y)$ are not physical.  As mentioned in
Section~\ref{sec:setup}, our choice of coordinates does not fix the
gauge completely. The residual gauge freedom appears numerically as
ripples in $B(t,\,y)$.  The precise form of these gauges modes is
worked out in Appendix~\ref{sec:residual_gauge}, and is in agreement
with the numerical plots.  The lowest frequency mode of these gauge
modes generically appears in the evolution of $B \left( t ,\, y
\right)\,$.  In the left panel of Figure~\ref{fig:ads_radion}, we see
this effect in the form of two bulk waves with period $2\,$, which
propagate on top of the profile of $B \left( t ,\, y \right)$. As
discussed in Section~\ref{sec:setup}, these gauge modes do not affect
the interbrane distance, as defined in equation~\eqref{d}.

 The numerical evolution of the interbrane distance for the 
AdS -- Schwarzschild solution is shown 
 in the right panel of
Figure~\ref{fig:ads_radion} for different values and signs of the
parameter $c\,$. For positive $c$ the branes approach each other,
while for negative $c$ they move apart.

\section{Instability of de Sitter Branes and 
Restructuring of Warped  Configurations}
 \label{sec:transition}

 Let us now study the evolution of the system (\ref{eq:eom}) in the
 presence of a bulk scalar field. We have to specify initial
 conditions, which do satisfy the constraint equations.  This task is
 now more complicated than it was without the scalar field.

\subsection{Instability of Warped Geometry with Curved Branes}
 \label{sec:inst}
 
 As a starting point we can check the code for known static warped
 geometry configurations (\ref{station}) with the scalar field and
 potentials chosen so as to stabilize the branes. We can then impose
 perturbations consistent with the constraint equations.  (This
 technique is described in detail in the Appendix
 ~\ref{sec:Perturbations})

 Static solutions of warped geometries with bulk scalar fields and
 with branes at the boundaries have been studied and classified
 in~\cite{FFK}.  In the 2d conformal gauge the static solutions with
 curved branes are given by
\begin{equation}\label{sta}
A(y,t)=B(y)+H \, t \,\,\, ,  \phi=\phi(y) ,
\end{equation}
where $B(y)$ and $\phi(y)$ are related through a set of ordinary
differential equations, which can be treated with the methods of
\cite{FFK}.

\begin{figure}[h]
  \centering
  \includegraphics[width=16.5cm,viewport=40 40 700 200,clip]{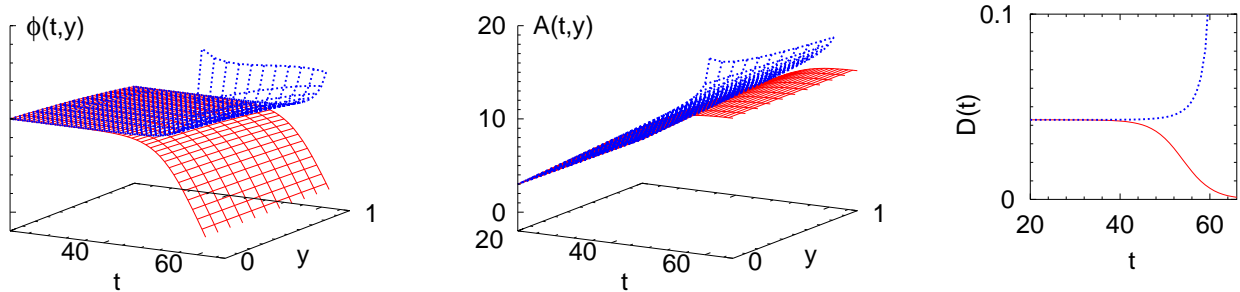}
  \caption{Instability of static solutions with dS branes: 
    Perturbations have induced significant departure from the static
    solution at $t\approx 40$.  The two unstable solutions shown
    correspond to positive (blue, upper surfaces and increasing $D$)
    and negative (red, lower surfaces and collapsing $D$) initial
    perturbations of $\delta \phi$, see
    Appendix~\ref{sec:Perturbations}.}
  \label{fig:instability}
\end{figure}

We use scalar field potentials (\ref{eq:pot}), which are designed for
brane stabilization.  The outputs of numerical integration of an
initially static configuration of two curved (de Sitter) branes and
bulk scalar filed with small perturbations around it is shown in
Figure~\ref{fig:instability}.
We see the
appearance of time-dependence in the initially
static field  $\phi(y)$, departure of metric
function $A$ from the hypersurface  described by 
equation $A(y,t)=B(y)+H \, t $,
and  a change in the interbrane distance $D$.
We show two  realizations of this model with different 
initial perturbations. From these results we conclude that, surprisingly, 
the static solutions with scalar field potentials
that  are supposed to stabilize branes
 are unstable for a range of $H$!

Fortunately, this unexpected result,
which we found here numerically,  can be independently obtained 
with analytical methods reported in the
accompanying paper \cite{tachyon}. Indeed, it is possible
to consider linearized perturbations of the bulk
scalar field $\delta \phi$ and scalar  metric perturbations
\begin{equation}\label{eq:metric:pert}
  ds^2 = W(y)^2 \left[(1+2\Phi) dy^2 + (1+2\Psi)
\left( -dt^2+e^{2Ht} d{\bf x}^2  \right) \right].
\end{equation} 
around the background warped geometry (\ref{warp}), where $\Phi$ and
$\Psi$ are small metric perturbations.  From the linearized Einstein
equations one can derive second order differential equations for the
fluctuations, which can be factorized into 4d massive scalar harmonics
on the de Sitter slices and KK eigenfunctions with eigenvalues $m$.
The lowest eigenvalue in the KK spectrum corresponds to the radion
mass.  The lowest eigenvalue $m^2$ is estimated as
\begin{equation}\label{crit1}
m^2=-4H^2+m_0^2(H) \ ,
\end{equation}
where $ m_0=\frac{2}{3} \, \frac{ \int \, d y \, {\rm e}^{-B}}{\int \,
  d y \, {\rm e}^{-B} {\phi'}^{-2} }$ is a functional of $H$.  In many
cases the first (negative) term in (\ref{crit}) exceeds the second
positive term, causing a tachyonic instability of the curved branes.
Indeed, the temporal part of the eigenmodes has an exponential
instability
\begin{equation}\label{asym}
  f_m (t) \propto e^{\mu \, t},
\end{equation}
where
\begin{equation}\label{mu}
\mu =\left( {\sqrt{\frac{9}{4}+\frac{|m^2|}{H^2}} - \frac{3}{2}} \right) \,  H.
\end{equation}
The time dependence from numerical calculations corresponding to
Figure~\ref{fig:instability} is consistent with the analytic result
(\ref{asym}).  In the limit of $H=0$ formula (\ref{crit1}) is reduced
to the known result for flat branes where the branes configuration is
stable \cite{stable}.  The curvature of the branes upsets the balance
between the bulk scalar gradients and its potentials, which otherwise
provide stabilization.

Thus, both from numerics and analytics we conclude that many static
configurations with de Sitter branes are unstable against classical
(or quantum) fluctuations.  While in the following we mostly discuss
the physical meaning and consequences of this result, here we also
note that this effect provides us with a tool to study brane dynamics
numerically. This is because we can start with a simple controllable
static configuration, without needing to resolve the time-dependent
constraint equations. Doing so, we can investigate numerically fully
non-linear time dependent dynamics due to the real physical tachyonic
instability of the initial configuration.

Depending on the sign of the initial perturbations (the coefficient
$\rho$ in equation \eqref{eq:gausspeak}) we encounter runaway behavior
towards smaller or larger interbrane distances as shown in
Figure~\ref{fig:instability}.  We consider this type of non-linear
dynamics in Section~\ref{sec:boom}.  Sometimes we do not find a
runaway behavior, but rather a re-structuring of brane configurations
as a transition between (at least) two static warped geometries. This
case will be considered in the next subsection.

\subsection{Dynamical Transition between Two Static Solutions}

As we discussed in the Introduction, the construction of brane models
with de Sitter branes is particularly challenging. Stable static solutions
with inflating branes can only be achieved provided the spatial
gradient of the bulk scalar field is sufficiently high, cf.\ equation \eqref{crit1}.

In the context of static warped geometries, brane embeddings can be
investigated in geometrical terms in a three dimensional phase space
\cite{FFK}.  This technique is especially useful to show that more
than one static solution for a given brane model, i.e.\ given
potentials $V(\phi)$ and $U_i(\phi)$, might exist as illustrated in
Figure~\ref{fig:phasespace}. Many of these solutions are unstable, as
shown above.  A fully numerical integration is a powerful (and maybe
the only) tool to study the non-linear dynamics of the unstable brane
configurations. A more comprehensive study of this issue, with
different bulk/brane potentials taken into account, will be presented
elsewhere. Here we limit our discussion to potentials of the
class~(\ref{eq:pot}).  In this subsection we discuss the case in which
the braneworld model admits one unstable and one stable static
solution, and the evolution of the system drives a transition between
the two.  Small perturbations around the unstable solution trigger the
tachyonic instability of the system, which is followed by a rapid
evolution of the bulk configuration.

An example of dynamical transition between two static brane
configurations is shown in Figure~\ref{fig:tuning}.  We plot the time
evolution of the distance $D(t)$ between the branes, the radion mass
$m^2$, the Hubble parameter (curvature) $H$, and the Weyl tensor
invariant $C^2$. The last two are defined as the averages of these
quantities over the extra dimension.  We observe a transition between
two states, from an initial brane configuration with higher brane
curvature (larger $H$) to a final configuration with lower curvature
(smaller $H$).  The first state is unstable; during this regime the
radion mass is tachyonic, $m^2 < 0$.  The value $H$ decreases with
time until the second term in \eqref{crit} dominates and the tachyonic
instability ceases.  In the cases we have studied, the decrease of $H$
is accompanied by a decrease of the physical interbrane distance,
until the stable configuration is reached.~\footnote{The quantity
  $H\,$ was defined in equation~(\ref{station}) only for static
  configurations.  During the time evolution, we choose to define it
  as the average over $y$ of $\: \dot B \left( t ,\, y \right) - \dot
  A \left( t,\, y \right) \:$ at any fixed time $t\,$.  In the
  examples discussed, we saw that the combination $\:A \left( t,\, y
  \right) - B \left( t,\, y \right)\:$ depended only weakly on $y$
  during the whole evolution.}  The final static configuration has
positive $m^2$.

The dynamics of the transition between the two static configurations
is quite violent, and is accompanied by a burst of the Weyl tensor
$C^2\,$.  The value of $C^2$ vanishes for the warped geometry
configurations at the beginning and the end of the transition.
\begin{subequations}
\begin{center}
\begin{figure}[h]
\begin{minipage}{16.5cm}
\begin{minipage}[b]{8cm}
\begin{flushleft}
   \includegraphics[width=8.25cm]{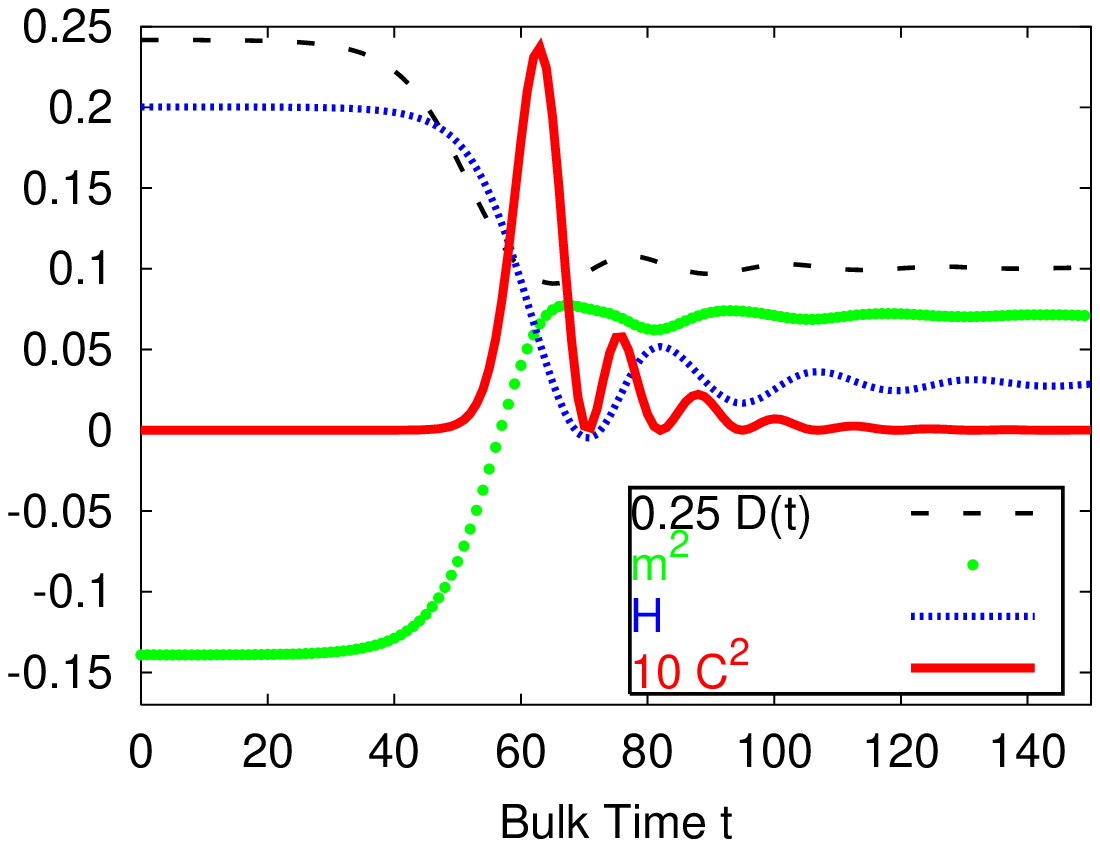}\\
\end{flushleft}
\par
\end{minipage}\hfill
\begin{minipage}[b]{8cm}
\begin{flushright}
  \includegraphics[width=8.25cm]{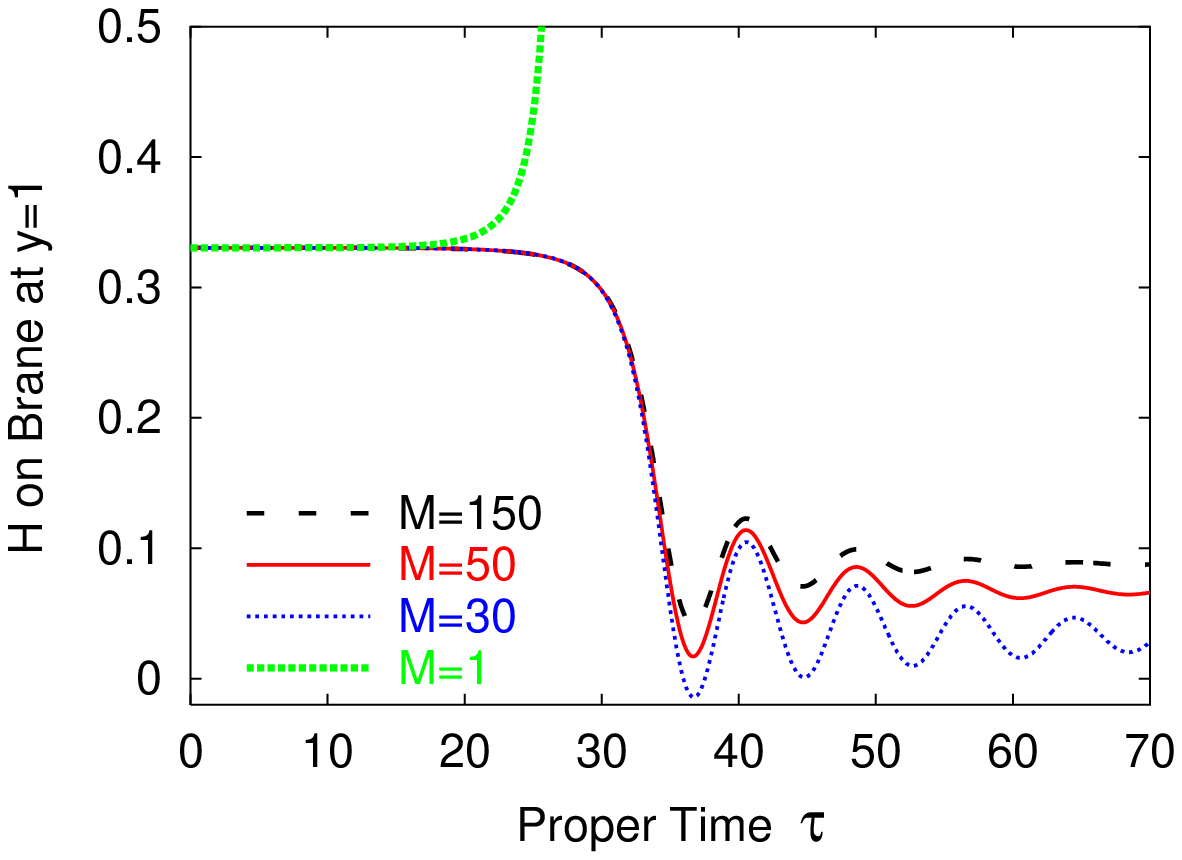}\\
\end{flushright}
\par
\end{minipage}
\caption{Left panel: Transition between two static brane
  configurations. Right panel: Transition observed on the brane at $y=1$
  for various values of the parameter $M\,$  of equation (\ref{eq:pot}), where the last plot
  $M=1$ corresponds to colliding branes (see Section~\ref{sec:boom}).}
\label{fig:tuning}
\end{minipage}
\end{figure}
\end{center}
\end{subequations}
Remember that we restrict ourselves to $(t, y)$ dependence and
``planar'' symmetry of the metric. Of course, the actual dynamics
between two warped configurations does not necessarily occur in this
class of metrics, and 3d inhomogeneities along the brane can be
excited.  As shown in \cite{tachyon}, the tachyonic instability of
warped geometry with de Sitter branes occurs for scalar
long-wavelength inhomogeneous modes with 3d momenta ${\bf k}$. The
present form of the {\tt BraneCode} cannot take them into account. We
assume that the background ${\bf k} \to 0$ mode dominates, but this
should be investigated in the future.  Tensor inhomogeneous modes do
not have tachyonic KK spectra \cite{FK} around the curved brane warped
geometry.  In fact, gravitational waves are absent for systems with
planar symmetry.  However, based on the evolution of the Weyl tensor
$C^2$, we conjecture that the actual dynamics is accompanied by a
burst of gravitational wave emission with 3d momenta of the order of
the non-adiabatic frequency $\sim 1/\Delta t$, where $ \Delta t$ is
the time of transition.  It will be interesting to check this with a
linear tensor mode analysis around the background geometry of the
Figure~\ref{fig:tuning}.

It is interesting to follow how the final state of the unstable
warped configuration depends on the parameters of the potentials.
We illustrate this with the parameter $M$ of the potential 
(\ref{eq:pot}).
In the example shown, the four dimensional cosmological evolution on
the two branes is characterized by a transition between two de Sitter
spaces.  In the right panel of
Figure~\ref{fig:tuning} we show how the four dimensional Hubble
parameter on one of the two branes changes as we change the brane
mass parameter $M$ of the scalar field, while leaving the
other parameters unchanged. In the limit of large $M$, the value
of the scalar field on the branes is always very close to its
expectation value $\sigma\,$. Moreover, phase space
portraits~\cite{FFK} (see Appendix~\ref{sec:shoot}) indicate
that the two static configurations approach each other in this
limit. This is also visible in Figure~\ref{fig:tuning}, where we see
that the difference between the initial and the final value of $H$
decreases as $M$ is increased. As an analogy, one may say that
higher mass parameters $M$ correspond to more rigid systems,
characterized by stiffer and quicker transitions between the two
static regimes. Tuning the parameters of the model, one can have flat Minkowski
branes in the stable final configuration.

In the  limit of negligible $M$ the system does not admit
stable configurations at all. The curve with $M = 1\,$, shown in
Figure~\ref{fig:tuning}, corresponds to a case in which the dynamics of
the system lead to a collision between the two branes. This case is
discussed in  detail in the next section.

\section{Brane Collisions}
\label{sec:boom}

Unstable warped configurations of curved (de Sitter) branes provide
suitable initial conditions for studying colliding branes, as we saw
in Subsection~\ref{sec:inst}. By controlling the initial fluctuations
(see Figure~\ref{fig:instability}) we can generate numerical runs 
with brane collisions.

The collision of branes is an interesting subject by itself.
In cosmology
colliding branes appear in models of
brane inflation \cite{infl} as well as in models without
(early universe) inflation \cite{coll1}. 
The latter models have difficulties 
which were discussed elsewhere (see e.g. \cite{coll2}).
In this paper, 
we focus on the issue of the bulk geometry and scalar field profiles
of
colliding brane configurations,  in a more general context.

In the  next Subsection ~\ref{sec:boom1}  we show a numerical example
of the brane collision and try to understand the properties
of the interbrane geometry. We find  that they become independent
 of the specific brane/bulk  potentials of the model.
In Subsection ~\ref{sec:kasner} we further argue that there is
a universal Kasner-like space-time asymptotic of the interbrane geometry.
This is a strong-gravity regime which can not be described in 4d by the moduli approximation.

\subsection{Geometry Between Branes}
\label{sec:boom1}

\begin{figure}[h]
\begin{minipage}{16.5cm}
\begin{minipage}[c]{10.5cm}
\begin{flushleft}
   \includegraphics[width=10.5cm,viewport= 50 35 360 235,clip]{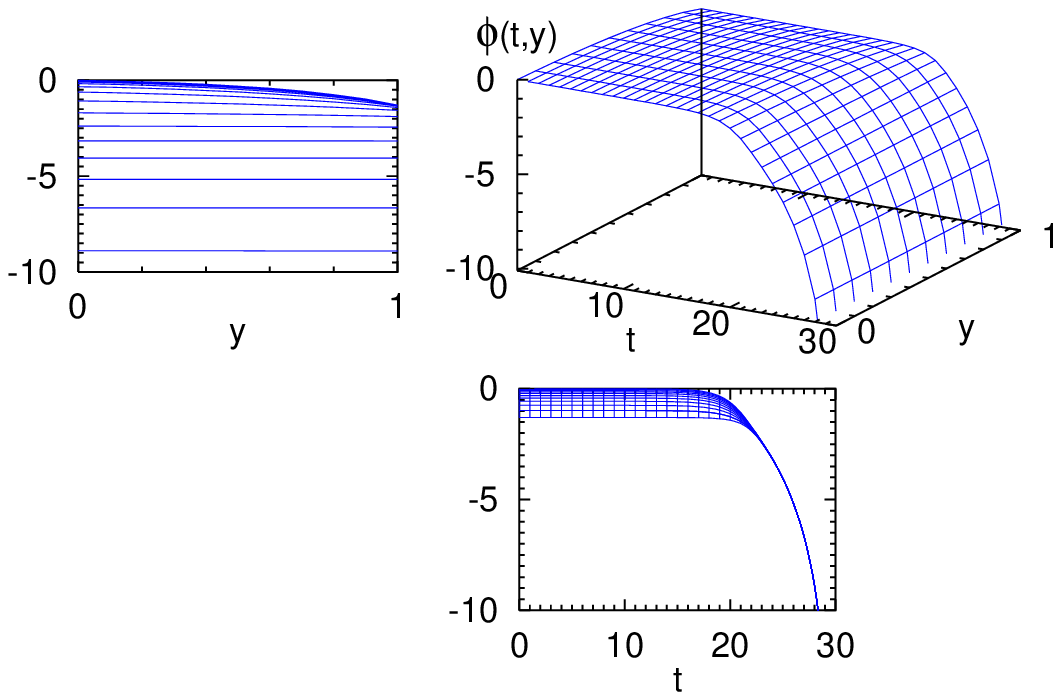}\\
\end{flushleft}
\par
\end{minipage}\hfill
\begin{minipage}[c]{6cm}
\begin{center}
  \includegraphics[width=6cm]{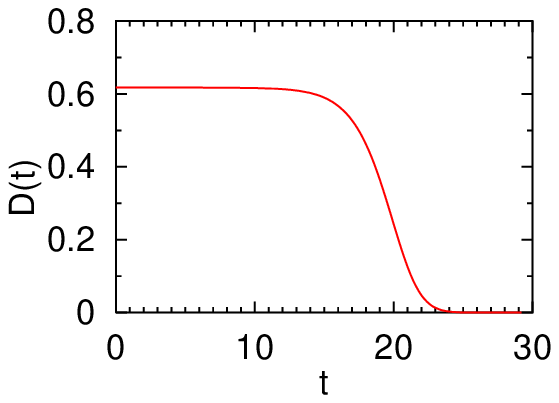}
\end{center}
\par
\end{minipage}
\end{minipage}
\caption{Flattening of $\phi$ gradients during the brane collision.
The left panel shows $\phi(y)$
for different
$t$, going from top to bottom. The lower panel shows $\phi(t)$ different $y$. In either plot you can see that $\phi$ becomes nearly homogeneous at later times.}
\label{fig:kasner_grad}
\end{figure}

 Figure~\ref{fig:kasner_grad} and Figure~\ref{fig:kasner_fit}
 show in detail the
evolution of the bulk scalar field $\phi(t,y)$, metric functions $A(t,y)$,
$B(t,y)$
and interbrane distance $D(t)$ in runs
which begin with an unstable warped de Sitter brane configuration and
end with a brane collision. 
%
The first thing to notice is that 
the system becomes homogeneous along the $y$ coordinate.
This is seen as the flattening of $\phi$ gradients over time. 
A similar flattening in $y$ direction occurs for the
metric components, see
Figure~\ref{fig:kasner_fit}.
 Also notice that the
absolute value of $\phi$ increases with time. This increase can be fit well by $\phi(t) \sim \ln t$.

A second feature of the brane collision is
the decrease of the metric component $e^B\,$; asymptotically
$B \rightarrow - \infty$ during the collision, cf.\ the
definition~(\ref{d}) of the interbrane distance. 
Recall that
the bulk/brane scalar field
potentials in the bulk equations (\ref{eq:eom}) and boundary
conditions  (\ref{eq:bc}) 
are
always multiplied by exponents $e^B$.
Therefore the contribution of 
the bulk/brane 
potentials becomes more and more negligible
in the dynamical equations (\ref{eq:eom})
 during the collision.
 
 This leads us to the important conclusion that asymptotically the
 dynamics of the brane collision do not depend on the form of the
 bulk/brane potentials.  Notice, however, a potential exclusion from
 this rule related to exponential potentials $e^{\alpha \phi}$.  In
 this case the typical logarithmic time divergence of $\phi$ leading
 up to the collision leads to the growth of the value of the
 potentials with time which may compensate the decrease of the metric
 function $e^B\,$.  In this paper we concentrate on the potentials
 (\ref{eq:pot}) where asymptotically the dynamics are potential-free.

This potential-free asymptotic immediately helps to explain heuristically
the first feature, why the system becomes homogeneous along the $y$ coordinate.
Indeed,  looking at the boundary conditions, we see that the
gradients of  $A\,$, $B\,$, and $\phi\,$ at the branes are
proportional to $\exp(B)$ and therefore vanish 
as $e^B \to 0$. 

Next, let us consider equations (\ref{eq:eom}) under the
assumption that the bulk/brane
potentials for the scalar field can be neglected
and that the geometry becomes homogeneous.
After this simplification equations~(\ref{eq:eom}) 
become ordinary differential equations, which
 can be easily solved.  We find
\begin{eqnarray}\label{eq:solboom2}
A &=&  A_0 + \frac{1}{3} \, {\rm ln }
\left( t_c - t  \right)
\,\,, \nonumber\\
\phi &=&  \phi_0 - r \,
{\rm ln } \left( t_c - t \right)
\,\,, \nonumber\\
B &=&  B_0-\frac{1}{3}\left(1-3\dot{B}_0 t_c-
\tfrac{3}{2} r^2\right)\frac{t}{t_c}-\frac{1}{3}\left(1-\tfrac{3}{2} r^2\right) \,
{\rm ln } \left( t_c - t \right)\ .
\end{eqnarray}
The constants of integration $A_0, B_0, \phi_0$ correspond to the
values of the fields at some time $t=0$.  The time $t=0$ cannot be the
beginning of integration where we know that the approximation does not
hold. We will give meaning to the integration constants shortly.  The
collision time is $t_c= -1/\dot A_0$.  We also introduce a convenient
intermediate parameter $r=\dot{\phi}_0 \, t_c$.  The brane collision
corresponds to values of $r$ satisfying $r^2 \leq \frac{2}{3}$.  The
scalar field potentials, as well as inhomogeneities along the $y$
coordinate, result in small corrections which are neglected
in~(\ref{eq:solboom2}).  Asymptotically the logarithmic terms in
(\ref{eq:solboom2}) dominate and we arrive at a homogeneous metric
with power-law dependence on time.  This is nothing but the
recognizable Kasner-like space-time metric.

\subsection{Universal Kasner-like Asymptotic}
\label{sec:kasner}

The regime when the logarithmic terms in (\ref{eq:solboom2}) determine
the behavior of the system corresponds to the universal Kasner
solution in five dimensions with a massless scalar field.  Four
dimensional homogeneous but anisotropic Kasner solutions with the
massless scalar field were constructed a long time ago in
\cite{Kasner}. Its higher dimensional generalization is obvious
\cite{cop}.  Indeed, in 5d we have the following exact solution with
the massless scalar field
\begin{align}
  \label{eq:kasner}
  ds^2&=-d\tau^2+\tau^{2p_\te{y}}dy^2+\sum\limits_{i=1}^3
  \tau^{2p_i}dx_i^2\,\,,\nonumber\\
   p_1+p_2+p_3+p_\te{y} &=1 \nonumber\\
  p_1^2+p_2^2+p_3^2+p_\te{y}^2 &=1-q^2  \nonumber\\
\phi&=q\ln \tau \, \ .
\end{align}
The vacuum Kasner solution has $q=0$. The parameter $q$ characterizes
the contribution of the scalar field.  The time $t$ in the 2d
conformal gauge (\ref{line}) and $\tau$ are related by transformation
\begin{equation} \label{kasner}
t = \tau^{1-p_\te{y}} .
\end{equation}
The significance of the Kasner-like space-time (\ref{eq:kasner})
is not only in the fact that it is an exact solution
of the Einstein equations, but mostly because it is a {\it generic}
asymptotic of {\it arbitrary} collapsing solutions \cite{BLK}.

In this section we explicitly demonstrate how the geometry of
colliding branes, as a case of the collapsing solution, approach the
universal Kasner-like asymptotic.

Kasner-like geometry as generic collapsing solution was already
advocated in string cosmology \cite{damour}.  As we show here, this
asymptotic also applies to brane cosmology (in other words string
cosmology with branes).  There is, however, a specific new feature
that appears in the brane cosmology case.  The isometry in the brane
directions is reflected in the additional constraint. In 5d
\begin{equation} \label{isom}
p_1=p_2=p_3 \ .
\end{equation}
This constraint and the two equalities for Kasner indices allow to
express $p_1$ and ${p_\te{y}}$ through the parameter $q$
\begin{equation}
 \label{eq:ind}
 p_1 =\frac{1}{4} \left(1 \pm \sqrt{1-\frac{4}{3} q^2} \right) \, , \hspace{1em}
{p_\te{y}} = \frac{1}{4} \left(1 \mp  3 \sqrt{1-\frac{4}{3} q^2} \right) \, .
\end{equation}
The range of the parameter $q$ is $-\frac{\sqrt{3}}{2} \leq q \leq
\frac{\sqrt{3}}{2}$; the range of $p_y$ and $p_1$ correspondingly are
$-\frac{1}{2} \leq p_y \leq 1$ and $0 \leq p_1 \leq \frac{1}{2}$.  In
the vacuum limit of vanishing $q$ one finds $p_1=0$ or $1/2$ and
$p_\te{y}=1$ or $-1/2$.

One of the feature of the Kasner-like asymptotic is a chaotic
alteration of the indices $p_A$ \cite{BLK}, with alternating
contraction and expansion in some of the directions. In the presence
of the scalar field the process ceases as all $p_A$ become positive.
For colliding branes $p_y > 0$, which gives $0 <p_1 < \frac{1}{3}$, so
both indices are positive and no alteration of indices is expected.

\begin{figure}[h]
  \centering
  \includegraphics[width=16.5cm]{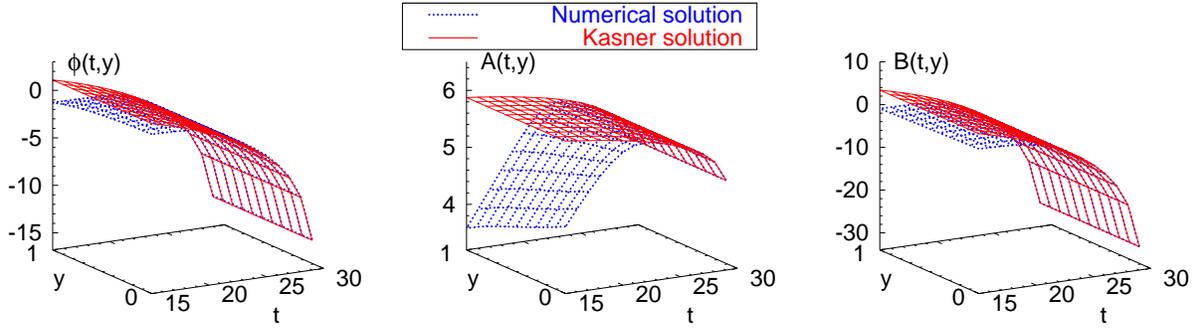}
  \caption{Numerical solutions (lower red surfaces) asymptotically approach universal
Kasner-like solution (upper blue  $y$ - independent surfaces).}
 \label{fig:kasner_fit}
\end{figure}

\begin{figure}[h]
  \centering
  \includegraphics[width=8.25cm]{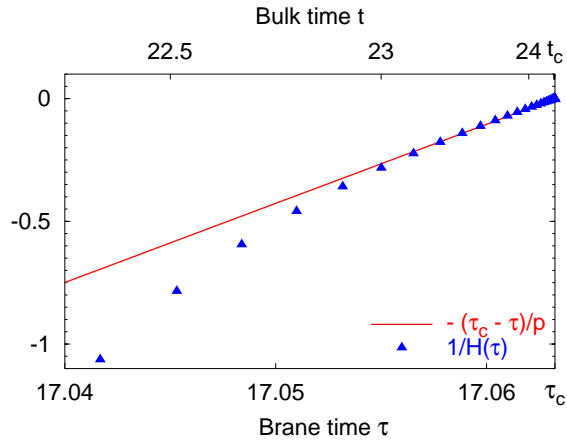}
  \caption{Induced Hubble parameter $H(\tau)$ at the  colliding branes}
  \label{fig:kasner_hubble}
\end{figure}

For our case (\ref{isom}) the Kasner solution (\ref{eq:kasner}) can be
rewritten to the 2d conformal gauge with the help of the time redefinition
(\ref{kasner})
\begin{equation}\label{eq:conf_kasner}
  ds^2=t^{\frac{2p_\te{y}}{1-p_\te{y}}}\left(dy^2-dt^2\right)+t^{\frac{1}{3}}d\vec
x^2\, .
\end{equation}
In terms of the metric functions $A$ and $B$ and the field $\phi$, the
Kasner solution (\ref{eq:conf_kasner}) reads as
\begin{eqnarray}
  \label{eq:collision_final}
  A&=&\frac{1}{3}\ln(t_c-t)\ ,\nonumber\\
  B&=&\frac{p_y}{1-p_y}\ln(t_c-t)\ ,\\
  \phi&=&\frac{q}{1-p_y}\ln(t_c-t)\ .\nonumber
\end{eqnarray}
The solution (\ref{eq:collision_final}) is identical to the leading terms of
(\ref{eq:solboom2}) by the identification $q=6r/(3r^2+4)$ 
%
%
Thus the integration constant $r$ in (\ref{eq:solboom2}) is related to
the parameters of the Kasner solution.
Figure~\ref{fig:kasner_fit} shows how the metric components $A$ and
$B$ and the field $\phi$ found numerically, approaches the universal
Kasner asymptotic (\ref{eq:conf_kasner}).
Next, consider  metrics  induced by the bulk Kasner geometry
 at  the branes (which is independent of $y$)
\begin{equation}
  \label{eq:kasner_induced}
  ds^2=-d\tau^2+(\tau_c-\tau)^{2 p_1} d\vec x^2\ ,
\end{equation}
The induced Hubble parameter on either branes is then given by
\begin{equation}
  \label{eq:kasner_hubble}
  H=-\frac{p_1}{t_c-t}\ .
\end{equation}
This time-dependence of the Hubble parameter at the brane is a good
fit to the asymptotic behavior of the Hubble parameter we found
numerically, as illustrated in Figure~\ref{fig:kasner_hubble}.
The induced metric at the brane {(\ref{eq:kasner_induced}) depends on
  the parameter $q$ through the index $p_1$.  This parameter is absent
  in the simple moduli approximation of the 4d effective theory at the
  brane, which does not take into account strong gravity arising in
  the bulk geometry.

\section{Summary}

We design the numerical code, {\bc}, to treat non-linear time-dependent
dynamics of 5d braneworlds with plane-parallel branes at the edge
and with bulk scalar field with arbitrary bulk and brane potentials.
It is possible to choose  a convenient gauge where the 
brane positions are not changing with time, and dynamics is imprinted in
two metric components and bulk
scalar field. These bulk equations for gravity and the scalar field
are supplemented by boundary conditions at the orbifold branes and
initial conditions in the bulk.
We also 
treat the constraint equations at the initial  time hypersurface.
So far we have only included a single bulk scalar field, but the
code could in principle be extended to include other layers such as
additional scalars in the bulk or on the branes.

We check the code for the brane models with known analytic solutions.
For two branes embedded in the 5d  background with negative 5d cosmological
constant  without the scalar field
we numerically reproduce generic AdS-Schwarzschild solution

Next, we consider more comprehensive braneworlds with the bulk scalar field.
We investigate numerically  small perturbations around warped 
stationary configuration with bulk scalar and with de Sitter branes
including the bulk/brane potentials, which are introduced 
for the brane stabilization.
However, for the large enough 4d curvature of inflating branes
the system is unstable and 
runs away from the initial warped configuration with de Sitter branes.
This effect is confirmed 
 independently  by an analytic calculation of small
scalar perturbations in this setting \cite{tachyon}. 
The scalar fluctuations around a warped
configuration with curved branes have as their lowest eigenvalue
\begin{equation}\label{crit}
m^2=-4H^2+m_0^2(H) \ .
\end{equation}
The term $m_0(H)$ is a functional of $H$, and depends
 on the parameters of the model.
If parameters are such that $m^2$ becomes negative due to
excessive curvature $\sim H^2$, the brane configuration
becomes unstable. For relatively low values of $H^2$
the radion mass (\ref{crit}) is positive and the system
is stable. 
Our   interpretation of this instability
is the following.
Stabilization of flat branes
 is based on the balance between
the gradient $\phi'$ of the
bulk scalar field and the brane  potentials $U(\phi)$
which tends to keep $\phi$ pinned down to its values $\phi_i$ at the branes.
The interplay between different forces becomes
 more delicate if the branes are curved,
and for the brane curvature exceeding some critical value
the brane configuration
becomes unstable.

Tachyonic instability of curved branes has serious implications for
theory of inflation in the braneworlds
and is discussed in details in the accompanying paper \cite{tachyon}.

Our numerical simulations allow us to follow dynamics of the brane configuration
triggered by the tachyonic instability.
The end point of evolution depends on the presence or absence of
another (or others) warped stationary configurations in the model.

The question about the multiplicity of warped solutions
can be studied in the framework
of warped geometry with no time-dependence.
We implemented the geometrical constructions in the phase space of
solutions
of the gravity plus scalar system that had been developed earlier. In
the model
with quadratic bulk and brane potentials, depending on the parameters
there are single or double warped solutions.

Thus we see that in some cases the warped branes system can admit
two solutions for the same parameters
of the potentials, with different values of the curvature of the de
Sitter brane, which is
proportional to $H^2$.
Suppose we start a numerical run with
 the warped solution that has a larger value of the brane curvature,
which is unstable. Then we observe
numerically that this configuration evolves dynamically and ends
up in the
the state which corresponds to the
second warped solution.
The second solution is stable if the corresponding radion mass $m^2$ is
positive,
as in the example shown in the text.
This re-structuring is accompanied by strong dynamical features
like a burst in the Weyl tensor,
which vanishes in the initial and final warped configurations.
Although inhomogeneous tensor modes are not included in the  code,
based on this behaviour of the Weyl tensor we conjecture that
brane re-structuring should be accompanied by the emission
of gravitational waves due to the non-adiabaticity of the process.

All together, this process looks like a decay of the meta-stable
state of the strongly curved branes due to the tachyonic instability
into the more stable state where the branes have lower curvature. This
transition is marked by a
burst of gravitational field anisotropy (gravitational waves?).
It will be interesting to investigate what applications this may have
to cosmology
with branes.
Another  potential application of brane restructuring would be the
problem of the 4d cosmological constant in the braneworld picture.
The cosmological constant problem was discussed recently from a braneworld
perspective, in which a low 4d cosmological constant corresponds to a
flat brane.
There was a
suggestion that the flat brane is a special solution of the bulk
gravity/dilaton system with a single brane \cite{const},
but the model has difficulties \cite{Nilles}.
In our setup, we consider two branes.
 The new element which
emerges from our paper
 is the instability of the curved branes.
So far we have only shown an example of re-structuring between
two curved brane configurations.
It will be interesting to see if there are brane models with
more than two stationary
warped geometry configurations, or with several scalar fields,
 and
 to investigate if there is a mechanism
for brane flattening.

Finally, we studied the geometry of colliding branes.
As initial conditions we used the unstable curved brane configurations
with parameters which do not allow another warped geometry
configuration.
In such cases the end point of the brane dynamics is either a brane
collision or branes moving apart.
We investigated in detail the geometry of colliding branes.
The bulk  metric and the bulk scalar field become homogeneous, i.e.
$y$-independent, and the brane dynamics
asymptotically does not depend on the scalar field potentials.
Instead, the geometry of colliding branes asymptotically approaches a
universal
Kasner-like solution with a free scalar field. It is known
that the
Kasner asymptotic is a generic solution of the high dimensional
gravity/dilaton system \cite{damour}.
In our case isometry of the brane slices
guarantees equality of the Kasner indices $p_1=p_2=p_3$.
In 5d this condition leaves only one single parameter
of the Kasner-like solution $q$, associated with the
bulk scalar field contribution. This parameter is determined by the
initial conditions. For the 5d brane system we considered,
there is no chaos in the alteration of the Kasner indices.
It will be interesting to investigate this issue for other
situations, for example when the form field is included and the brane dimensions
and co-dimensions
are different.

\section*{Acknowledgments}

We are grateful to 
A. Linde, S. Mukohyama, A. Peet and D. Pogosyan  for valuable discussions.
 This research was supported in part
by the Natural Sciences and Engineering Research Council of Canada and
CIAR.

\appendix
\section*{Appendices}

\section{Choice of Gauge}

The system we are studying has a gauge freedom which amounts to
different possible coordinates for the five dimensional metric and for
the positions of the two branes. Our choice not only aims at
simplifying the equation of motions, but also at removing
``redundant'' degrees of freedom which would not allow us to write a
closed system of equations for the numerical integration. In
Section~\ref{sec:setup}, we claimed that it is always possible to
choose a system of coordinates in which the $(t,y)-$part of the metric
is conformally flat and in which the two branes are fixed at the
positions $y=0$ and $y=1\,$, irrespective whether their physical
separation is constant or changing in time. We show this explicitly in
Subsection~\ref{sec:comotion}.  This choice does not fix the gauge
completely, however.  The form of the remaining gauge degrees of
freedom, which are expected to affect the numerical solutions, is
worked out in Subsection~\ref{sec:residual_gauge}.

\subsection{Comoving Branes}
\label{sec:comotion}

By assumption, the system is homogeneous and isotropic along the
spatial coordinates ${\bf x}\,$, and the position of each brane in the
extra space is specified by a function of time only (parallel
branes). Since the metric coefficients depend only on the two
coordinates $t$ and $y\,$, the metric can be written in the 2d
conformal gauge \eqref{line}

The change of coordinates
\begin{eqnarray}
t &\rightarrow& {\bar t} = \frac{f \left( t + y \right) + g \left( t -
y \right)}{2} \,\,, \nonumber\\
y &\rightarrow& {\bar y} = \frac{f \left( t + y \right) - g \left( t -
y \right)}{2} \,\,,
\label{free}
\end{eqnarray}
where $f$ and $g$ are two arbitrary scalar functions, preserves the
2d conformal gauge, since it affects the metric (\ref{line}) only by the
change
\begin{equation}
B \left( t , y \right) \rightarrow {\bar B} \left( {\bar t}, {\bar y}
\right) = \left[ B \left( t, y \right) - \frac{1}{2} {\rm ln} \left(
f' \left( t + y \right) g' \left( t - y \right) \right) \right]_{t,y
\rightarrow {\bar t}, {\bar y}}\ .
\end{equation}
This is most easily seen in null coordinates, where
\begin{equation}
t = \frac{v + u}{\sqrt{2}} \,\,,\,\,
y = \frac{v - u}{\sqrt{2}} \;\;\Rightarrow\;\; ds^2 = - 2 \, du \, dv
\,\,.
\end{equation}

Generally, the two branes will have a time dependent position in the
extra dimension, described by the two functions $y_1 \left( t \right)$ and
$y_2 \left( t \right) \,$, respectively. However, as long as their motion
occurs slower than the speed of light, $\vert \dot{y}_1 \vert$ and
$\vert \dot{y}_2 \vert < 1\,$, we can perform a change of
coordinates~(\ref{free}) to have them at fixed position along $y\,$,
as we now show.

Let us first fix the first brane at $ y \equiv 0\,$. For this to
happen, the two functions $f$ and $g$ appearing in equations~(\ref{free})
have to satisfy
\begin{equation}
f \left( t + y_1 \left( t \right) \right) = g \left( t - y_1 \left( t
\right) \right) \,\,.
\label{zerofix}
\end{equation}
We can choose $f$ arbitrarily, and use~(\ref{zerofix}) to determine
$g\,$. The condition $\vert \dot{y}_1 \vert <1$ guarantees this can be
always done, since the arguments of both the two functions increase
monotonically in time.

In the new coordinate system, the first brane is fixed at $y_1 \equiv
0\,$, while again the second one will be generally moving according to some
function ${\tilde y}_2 \left( t \right)\,$. This function describes
the parallel motion of the second brane with respect to the first
one. Since in the old system of coordinates the relative motion was at
a speed lower than that of light, this will be the case also in the new
coordinates, $\vert \dot{\tilde y}_2 \vert < 1 \,$. To preserve the
first brane at the origin, the residual freedom (\ref{free}) is
restricted to $f \lmk w \rmk = g \lmk w \rmk\,$, i.e.\ $f$ and $g$
are the same functions of their arguments. If we choose $f$ to
satisfy
\begin{equation}
f \lmk t + {\tilde y}_2 \lmk t \rmk \rmk = f \lmk t - {\tilde y}_2
\lmk t \rmk \rmk + 2 \,\,,
\label{rest}
\end{equation}
we finally reach a third system of coordinates where the two branes
are fixed at $y_1 \equiv 0$ and $y_2 \equiv 1\,$, respectively. As
before, the function $f$ can always be constructed. Since $\vert
\dot{\tilde y}_2 \vert < 1\,$, the arguments of both terms increase
monotonically in time. We can then use the value of $f$ at the right
hand side of equation~(\ref{rest}) to ``construct'' the value of $f$ at the
left hand side.

\subsection{Residual Gauge Freedom}
\label{sec:residual_gauge}

Even with the position of the branes fixed, the freedom of
reparametrization is not exhausted yet. The residual gauge degrees of
freedom are again of the form (\ref{free}), with
\begin{equation}
\label{eq:fg2}
f \lmk w \rmk = g \lmk w \rmk \equiv F \lmk w \rmk 
\quad\quad, \quad F \lmk w + 2 \rmk - F
\lmk w \rmk = 2 \,\,.
\end{equation}
The most generic function $F$ with this property is
\begin{equation}
  \label{eq:residual_freedom}
  F(w)= w +\sum\limits_{n=0}^{\infty}\lmk a_n\cos n\pi w+b_n
  \sin n \pi w\rmk \,\,,
\end{equation}
with arbitrary coefficients $a_n, b_n\,$.  The appearance of these
gauge degrees of freedom is manifest in some of the numerical results
we obtained, for example in Figure~\ref{fig:ads_radion} they are shown
as ripples in the metric component $B(t,y)$.

\subsection{Perturbations of the Randall-Sundrum geometry}
\label{sec:rspert}

It is interesting to note that, apart from pure gauge modes described
in the previous appendix, there exists only one kind of ${\bf
x}-$independent small fluctuations about the Randall-Sundrum
geometry~(\ref{eq:rs_sol}).  As we show now, this perturbation is
related to a small change of the interbrane distance $D\,$, which is
not stabilized without a scalar field.  We know
that any ${\bf x}-$independent configuration can be written in the
conformal gauge with the position of the branes at $y=0,1\,$. Thus, all the
perturbations we are interested in can be written in terms of the
metric components $A \left( t, y \right) = A_0(y) + \delta a \left( t, y
\right)\,$, $B \left( t, y \right) = B_0(y) + \delta b \left( t, y
\right)\,$, where $A_0=B_0$ is the Randall--Sundrum
solution~(\ref{eq:rs_sol}). To find which perturbations are allowed, we linearize the
bulk Einstein equations. The dynamical ones reduce to
\begin{eqnarray}
&& \ddot {\delta a} - \delta a'' + \frac{6 \delta a'}{y + \gamma} +
\frac{8 \delta b}{\left( y + \gamma\right)^2}  = 0\,\,,  \nonumber\\
&& \ddot {\delta b} - \delta b'' - \frac{6 \delta a'}{y + \gamma} -
\frac{4 \delta b}{\left( y + \gamma\right)^2} = 0 \,\,,
\label{lindyn}
\end{eqnarray}
while the two constraint equations are conveniently recast in the form
\begin{eqnarray}
\frac{d}{d t}\left( \delta a' + \frac{\delta b}{y + \gamma}
\right) &=& 0\,\,, \nonumber\\
\frac{d}{d y}\left[ \left( y + \gamma \right)^{-3} \left( \delta
a' + \frac{\delta b}{y + \gamma} \right) \right]&=& 0 \,\,.
\end{eqnarray}
Here $\gamma = \left( {\rm e}^{D/l} - 1 \right)^{-1}\,$, cf.\
equation~(\ref{eq:rs_sol}). The last two equations give
\begin{equation}
\delta a' +  \frac{\delta b}{y + \gamma} \equiv C \left( y + \gamma \right)^3\,\,,
\label{linco}
\end{equation}
with $C$ constant.
Finally, linearizing the boundary conditions we have
\begin{equation}
\left( \delta a' + \frac{\delta b}{y + \gamma} \right)_{\vert y=0,1} =
0 \quad,\quad \left( \delta b' + \frac{\delta b}{y + \gamma}
\right)_{\vert y=0,1} = 0 \,\,,
\label{linbc}
\end{equation}
the first of which enforces $C=0\,$. Substituting equation~(\ref{linco})
into the second equation of~(\ref{lindyn}), we have a differential equation in
terms of $\delta b$ and its derivatives only. Fourier transforming
\begin{equation}
\delta b = \int d \omega \, {\rm e}^{i \,
\omega\,t} \, {\widetilde {\delta b}} \left( \omega, y \right)\,\,,
\label{four}
\end{equation} 
we get an ordinary differential equation which is solved by
\begin{equation}
{\widetilde {\delta b}} = F \left[ {\rm cos } \left( \omega \, z \right) -
\frac{{\rm sin } \left( \omega \, z \right)}{ \omega \, z}  \right] 
+ G \left[ {\rm sin } \left( \omega \, z \right) +
\frac{{\rm cos } \left( \omega \, z \right)}{ \omega \, z}  \right]
\quad\quad,\quad z = y + \gamma \,\,.
\label{delb}
\end{equation}

The boundary conditions for $\delta b$ at the two branes become two equations for the
two parameters $F$ and $G\,$. Non-vanishing solutions are possible only for
\begin{equation}
\omega = n \, \pi \quad, \quad \quad n = 0, 1, 2, \dots \,\,.
\end{equation}
For these values, the two coefficients $F$ and $G$ are related by
$G = F \, {\rm tan } \, \omega \, \gamma \,$. The Fourier transform of $\delta a$
is then easily obtained from the remaining two equations
\begin{equation}
\delta a = \frac{1}{\omega \, z} \, \left[ - F \, {\rm sin } \left(
\omega \, z \right) + G \, {\rm cos } \left( \omega z \right) \right]
+ K \delta \left( \omega \right)
\,\,,
\end{equation}
where $K$ is a constant. Back in coordinate space
\begin{eqnarray}
\delta a &=& \Sigma_n \frac{- \, F_n}
{{\rm cos }\left( n \pi \gamma \right)} \,
\frac{{\rm e}^{i \, n \, \pi \, t} \,
{\rm sin } \left( n \, \pi \, y \right)}
{n \, \pi \, \left( y + \gamma \right)} + K\,\,, \nonumber\\
\delta b &=& \Sigma_n \frac{F_n}
{{\rm cos } \left( n \pi \gamma \right)} \,
{\rm e}^{i \, n \, \pi \, t} \left[ 
{\rm cos } \left( n \, \pi \, y \right)
- \frac{{\rm sin } \left( n \, \pi \, y \right)}
{n \, \pi \, \left( y + \gamma \right)}
\right]\,\,.
\label{linsol}
\end{eqnarray}

These are the most generic ${\bf x}-$independent perturbations of the Randall-Sundrum
solution~(\ref{eq:rs_sol}). However, most of them are pure gauge modes. Let us consider
infinitesimal change of coordinates of the residual gauge discussed
in Appendix~\ref{sec:residual_gauge}
\begin{eqnarray}
F \left( w \right) &=& w + {\tilde F} \left( w \right) = w + \sum_n f_n {\rm e }^{i \, n \, \pi \, w}
\,\,, \nonumber\\
t &\rightarrow& t + \sum_n f_n {\rm e }^{i \, n \, \pi \, t} \, {\rm cos } \left( n \, \pi \, y \right)
\,\,, \nonumber\\
y &\rightarrow& y + \sum_n f_n {\rm e }^{i \, n \, \pi \, t} \, i \, {\rm sin } \left( n \, \pi \, y \right)
\,\,.
\label{eq:linchange}
\end{eqnarray}
Under this infinitesimal change of coordinates, the metric coefficients undergo the infinitesimal changes
\begin{eqnarray}
A_0 &\rightarrow& A_0 + A_0' \, \frac{{\tilde F \left( t + y \right)} - {\tilde F \left( t - y \right)}}{2} \,\,,\nonumber\\
B_0 &\rightarrow& B_0 - B_0' \, \frac{{\tilde F \left( t + y \right)} - {\tilde F \left( t - y \right)}}{2} +
 \frac{{\tilde F' \left( t + y \right)} + {\tilde F' \left( t - y \right)}}{2}
\,\,.
\end{eqnarray}
By choosing
\begin{equation}
f_n = \frac{i \, F_n}{n \, \pi \, {\rm cos } \left( n \, \pi \, \gamma \right)}
\quad\quad,\quad n = 1,2,\dots\,\,,
\end{equation}
we see that the perturbations~(\ref{linsol}) are equivalent to
\begin{eqnarray}
A_0 + \delta a &\equiv& A_0 - \frac{F_0 \, y}{y + \gamma} + K = A_0 + \frac{F_0 \, \gamma}{y + \gamma} + {\tilde K}
\,\,\nonumber\\
B_0 + \delta b &\equiv& B_0 + \frac{F_0 \, \gamma}{y + \gamma} \,\,,
\end{eqnarray}
where ${\cal K} = K - F_0$ is also constant.

By an appropriate rescaling of the spatial ${\bf x}$ coordinates we can set
$\tilde K$ to zero.


\section{Determination of Static Configurations}
\label{sec:shoot}

Here we describe the method used to determine static solutions, once
the bulk and brane potentials for the scalar field are given. This is
done by a numerical boundary value problem solver using the shooting
method. As discussed in Section~\ref{sec:static}, we first deal
with the two boundary conditions at the first brane. Any two of $B_0
,\, \phi_0 ,\,\phi_0'\,$, and $H\,$ can be chosen freely, while the
other two are determined by the junction conditions at the first
brane. We find it more convenient to choose the values of $B_0$ and
$\phi_0'\,$, since the latter cannot be taken arbitrarily large if we
want the solutions to remain regular all across the bulk. A fourth
order Runge-Kutta integrator is then employed to integrate
equations~(\ref{eq:static}) in the bulk. The aim is to find the values
of $B_0$ and $\phi_0'$ for which the solutions are regular in the $0
\leq y \leq 1$ interval, and for which the boundary conditions on the
second brane are also satisfied. We can recast the latter in the form

\begin{equation}
c_1 \equiv B_1' - \frac{1}{6} {U}_1 \, {\rm e}^{B_1} = 0
\quad\quad,\quad c_2 \equiv \phi'_1 + \frac{1}{2} \frac{d {U}_1}{d \phi} \, {\rm e}^{B_1} = 0 \,\,.
\end{equation}

Both $c_1$ and $c_2$ are only functions of the chosen value for $B_0$
and $\phi_0'\,$, and in general do not vanish. We use Newton's method
to find the zeros of these two functions, that is the initial
configurations at the first brane for which the junction conditions at
the second brane are also satisfied. In practice, for the potentials
we have studied, Newton's method does not converge
globally. Fortunately, the bulk equations~(\ref{eq:static}) can be
integrated very quickly, so that we can perform a ``brute force'' scan
in the $\left\{ B_0, \phi'_0 \right\}\,$ plane. We then apply Newton's
method starting only from those values which are sufficiently close to
a solution, i.e.\ for which $c_1$ and $c_2\,$ turn out to be sufficiently
close to zero.

\begin{figure}[h]
\centering\includegraphics[width=11cm]{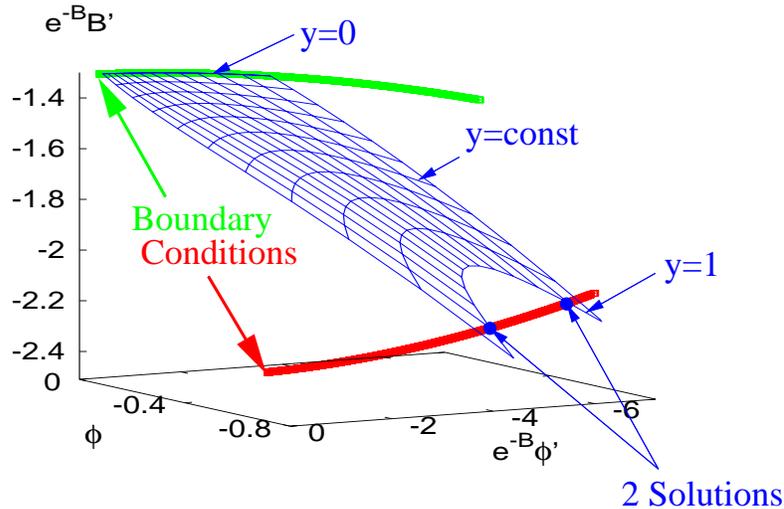}
\caption{Phasespace illustration of the two solutions for
  given potentials}
\label{fig:phasespace}
\end{figure}

The existence of static solutions is not guaranteed for arbitrary
bulk and brane potentials. As discussed in~\cite{FFK}, many potentials
do not give static solutions at all, while some others typically
lead to a finite number of them. Using the geometrical method of phase
portraits for quadratic potentials~(\ref{eq:pot}) we found at most two
static solutions in the (wide) space of possible initial
configurations we have scanned. Figure~\ref{fig:phasespace} shows how
the phase portrait method allows us to visualize the quest for
static configurations. Following the method of~\cite{FFK}, we draw
curves in the $\{\phi,e^{-B}\phi',e^{-B}B'\}$ phase space.

Each of the two thick curves refers to one of the two branes, and it
joins points for which the junction conditions on that brane are
satisfied. The thin curves are are a sampling of bulk trajectories
which satisfy the junction conditions at one of the two branes (at
$y=0$). Valid static solutions consist of trajectories that satisfy
the junction conditions at both branes. Hence, in the phase portrait
they are represented by the few trajectories which intersect both the
thick curves. Lines perpendicular to the trajectories represent lines
of $y=const$. We see that in the case at hand, corresponding to
quadratic potentials~(\ref{eq:pot}), there are two intersections.

\subsection{Perturbations of Static Solutions}
\label{sec:Perturbations}

Generic perturbations around a static configuration~(\ref{station})
are described by the functions $\delta B_0 \left(y \right)$, $\delta
A_0 \left(y \right)$, $\delta \phi_0 \left(y \right)$ and their first
time derivatives $\delta \dot B_0 \left(y \right)$, $\delta \dot A_0
\left(y \right)$, and $\delta \dot \phi_0 \left(y \right)$, which are
obtained by equating the time dependent fields and the first time
derivatives at an initial moment $t_0=0$. In the bulk, four of them
can be specified arbitrarily, while the remaining functions are
obtained from the constraint equations. One possible choice of initial
perturbations, adopted in the example of Figure~\ref{fig:instability},
is given by
\begin{eqnarray}
\label{eq:perturbations}
\delta B_0 \left(y \right) &=& \delta A_0 \left(y \right) = \delta
\dot A_0 \left(y \right) \equiv 0 \,\,, \nonumber\\
\delta \phi_0 \left(y \right) &\equiv& \delta \phi \left(y \right) \,\,,
\nonumber\\
\delta \dot \phi_0 \left(y \right) &=& \frac{H}{A^\prime} \,
\phi^\prime-\left(\frac{H^2}{{A^\prime}^2}{\phi^\prime}^2 +
\delta {\phi^\prime}^2 - 2 \, \phi^\prime \, \delta \phi^\prime - 2 \,
e^{2 B} \lkk V \left( \phi \right) - V \left( \phi - \delta \phi
\right) \rkk\right)^{\frac{1}{2}} \,\,, \nonumber\\
\delta\dot B_0 \left( y \right) &=& \frac{1}{3} \,
\frac{\phi^\prime}{A^\prime} \, \delta \dot \phi \,\,.
\end{eqnarray}

The perturbation $\delta \phi \left( y \right)$ can be specified
arbitrarily. In the example shown, the Gaussian profile
\begin{equation}
\label{eq:gausspeak}
\delta \phi = \rho\, {\rm exp } \left( -\,\frac{\left(
y-1/2 \right)^2}{2\,h^2} \right)\,\,,
\end{equation}
is centered between the branes. A sufficiently small values for $h$ guarantee that the
perturbations are exponentially suppressed at the brane locations,
leaving the junction conditions (practically) unaffected.

\begin{sidewaystable}[h!]\footnotesize
  \centering
  \begin{tabular}[]{|c|c|c|l c|c|c|c|}
    \hline
     &Figure \ref{fig:rs} &Figure \ref{fig:dark_radiation},\ref{fig:ads_radion}& &\hspace{\shift}Figure \ref{fig:instability} &Figure \ref{fig:tuning} &Figure \ref{fig:kasner_grad},\ref{fig:kasner_fit},\ref{fig:kasner_hubble}&Figure \ref{fig:phasespace} \\
     \hline
     \pb{\cwl}{%
       {\bf Bulk:}\\
       $V=\tfrac{1}{2}m^2\phi^2+\Lambda$
     }
     & 
     \pb{\cws}{%
       \begin{tabular}{l r}
         \pb{\cwh}{\begin{align*}
           m&=0\\
           \Lambda&=-6
         \end{align*}}
       &
       \pb{\cwh}{
         \begin{align*}
           m&=0.5\\
           \Lambda&=-6
         \end{align*}}
       \end{tabular}
     }
     &
     \pb{\cw}{%
       \begin{align*}
         m&=0\\
         \Lambda&=-6
       \end{align*}
     }
     &
     \pb{0.2cm}{%
         \begin{align*}
           m&=\\
           \Lambda&=
         \end{align*}
       }
     &
     \hspace{\shift}\pb{\cw}{%
       \begin{align*}
         0.5\\
         -6
       \end{align*}
     }
     &
     \pb{\cw}{%
       \begin{align*}
         0.5\\
         -6
       \end{align*}
     }
     &
     \pb{\cw}{%
       \begin{align*}
         0.5\\
        -6
       \end{align*}
     }
     &
     \pb{\cw}{%
       \begin{align*}
         1\\
         -6
       \end{align*}
     }
     \\
     \hline
     \pb{\cwl}{%
       {\bf Branes:}\\
       $U_i=$\\$\tfrac{1}{2}M_i(\phi-\sigma_i)^2+\lambda_i$
     }
     & 
     \pb{\cws}{%
       \begin{tabular}{l r}
         \pb{\cwh}{\begin{align*}
           M_0&=0\\
           \lambda_0&=6\\
           \\
           M_1&=0\\
           \lambda_1&=-6\\
         \end{align*}}
       &
       \pb{\cwh}{
         \begin{align*}
           M_0&=300\\
           \lambda_0&=5.98\\
           \sigma_0&=0.5\\
           M_1&=300\\
           \lambda_1&=-6.04\\
           \sigma_1&=0.60
         \end{align*}}
       \end{tabular}
     }
     &
     \pb{\cw}{%
       \begin{align*}
         M_i&=0\\
         \lambda_0&=6\\
         \lambda_1&=-6
       \end{align*}
     }
     &
     \pb{0.2cm}{%
         \begin{align*}
           M_0&=\\
           \lambda_0&=\\
           \sigma_0&=\\
           M_1&=\\
           \lambda_1&=\\
           \sigma_1&=
         \end{align*}
       }
     &
     \hspace{\shift}\pb{\cw}{%
       \begin{align*}
         2\\
         36.6\\
         0.201\\
         2\\
         -49.7\\
         -0.692
       \end{align*}
     }
     &
     \pb{\cw}{%
       \begin{align*}
         300\\
         6.04\\
         0.000202\\
         300\\
         -7.02\\
         -0.406
       \end{align*}
     }
     &
     \pb{\cw}{%
       \begin{align*}
         2\\
         6.24\\
         0.33\\
         2\\
         -102\\
         -1.05
       \end{align*}
     }
     &
     \pb{\cw}{%
       \begin{align*}
         100\\
         7.86\\
         0.00815\\
         100\\
         -13.9\\
         -0.715
       \end{align*}
     }
     \\
    \hline
     \pb{\cwl}{%
       {\bf Initial conditions for static solution:}
     }
     & 
     \pb{\cws}{%
       \begin{tabular}{l r}
         \pb{\cwh}{\begin{align*}
             H&=0\\
             \phi_0&=0\\
             \phi_0^\prime&=0\\
             B_0&=0\\
             B_0^\prime&=-1
           \end{align*}}
           &
           \pb{\cwh}{
             \begin{align*}
               H&=10^{-4}\\
               \phi_0&=0.5\\
               \phi_0^\prime&=0\\
               B_0&=0\\
               B_0^\prime&=-0.997
             \end{align*}}
       \end{tabular}
     }
     &
     \pb{2cm}{%
       (cf.\ Sec.\ \ref{sec:ads-schw})
       \centering 
       \vspace{-0.3cm}\begin{align*}
         H&=0\\
         D&=\ln 2\\
         c&=1\\
         \dot A_0&=y+1\\
         \dot B_0&=-\dot A_0
       \end{align*}
     }
     &
     \pb{0.2cm}{%
         \begin{align*}
           H&=\\
           \phi_0&=\\
           \phi_0^\prime&=\\
           B_0&=\\
           B_0^\prime&=
         \end{align*}
       }
     &
     \hspace{\shift}\pb{\cw}{%
       \begin{align*}
         0.3\\
         0\\
         -0.02\\
         -0.3\\
         -0.304
       \end{align*}
     }
     &
     \pb{\cw}{%
       \begin{align*}
         0.2\\
         0\\
         -0.05\\
         0.5\\
         -1.66
       \end{align*}
     }
     &
     \pb{\cw}{%
       \begin{align*}
         0.3\\
         0\\
         -0.3\\
         -0.1\\
         -0.96
       \end{align*}
     }
     &
     \pb{\cw}{%
       \begin{align*}
         0.0354&&0.368\\
         -0.0508&&-0.00276\\
         -0.393&&-0.241\\
         -2.01&&-0.818\\
         -0.179&&-0.579
       \end{align*}
     }
     \\
     \hline
     \pb{\cwl}{%
       {\bf Mass bound:}\\
     }
     & 
     \pb{\cws}{%
       \begin{tabular}{l r}
         \pb{\cwh}{\begin{align*}
           m_\te{eff}^2&=0
         \end{align*}}
       &
       \pb{\cwh}{
         \begin{align*}
           m_\te{eff}^2&<0.0016
         \end{align*}}
       \end{tabular}
     }
     &
     \pb{2cm}{%
       no static solution
     }
     &
     \pb{0.2cm}{%
         \begin{align*}
           m_\te{eff}^2&<
         \end{align*}
       }
     &
     \hspace{\shift}\pb{\cw}{%
       \begin{align*}
         -0.360
       \end{align*}
     }
     &
     \pb{\cw}{%
       \begin{align*}
         -0.139
       \end{align*}
     }
     &
     \pb{\cw}{%
       \begin{align*}
         0.041
       \end{align*}
     }
     &
     \pb{\cw}{%
       \begin{align*}
         0.17&&-0.387
       \end{align*}
     }
     \\
     \hline
     \pb{\cwl}{%
            {\bf Perturbations:}\\
     }
     & 
     \pb{\cws}{%
       \flushleft
       no
     }
     &
     \pb{\cw}{%
       no
     }
     &
     &
     \hspace{\shift}\pb{\cw}{%
       \begin{align*}
         \delta\phi&=c_1e^{c_2(y-\tfrac{1}{2})^2}\\
         c_1&=\pm 10^{-6}\\
         c_2&=-1
       \end{align*}
     }
     &
     \pb{\cw}{%
       no
     }
     &
     \pb{\cw}{%
       no
     }
     &
     \pb{\cw}{%
       no
     }
     \\
     \hline
  \end{tabular}
  \caption{Parameters used for simulations presented in the figures}
  \label{tab:parameters}
\end{sidewaystable}
\clearpage

\providecommand{\href}[2]{#2}\begingroup\raggedright

\end{document}